\documentclass[twocolumn,aps,floatfix,superscriptaddress,pra,nobibnotes,longbibliography]{revtex4-1}

\usepackage{graphicx}
\usepackage{amsmath,amsthm,amssymb,amsfonts}

\usepackage[USenglish]{babel}
\usepackage[bbgreekl]{mathbbol}
\usepackage{color}

\usepackage[colorlinks=true,linkcolor=blue,citecolor=blue]{hyperref}
\usepackage{accents}
\newlength{\dhatheight}
\newcommand{\doublehat}[1]{%
    \settoheight{\dhatheight}{\ensuremath{\hat{#1}}}%
    \addtolength{\dhatheight}{-0.35ex}%
    \hat{\vphantom{\rule{1pt}{\dhatheight}}%
    \smash{\hat{#1}}}}

\definecolor{cg}{rgb}{0.0, 0.5, 0.0}

\def\d{\mathrm{d}}
\def\rmi{\mathrm{i}}

\def\L{\mathcal{L}}
\newcommand{\udln}[1]{%
    \underline{#1}}

\begin{document}

\title{Density-wave steady-state phase of dissipative ultracold fermions with nearest-neighbor interactions}

\author{Jaromir Panas}
\email{panas@th.physik.uni-frankfurt.de}
\affiliation{Institut f\"ur Theoretische Physik, Goethe-Universit\"at, 60438 Frankfurt am Main, Germany}

\author{Michael Pasek}
\affiliation{Institut f\"ur Theoretische Physik, Goethe-Universit\"at, 60438 Frankfurt am Main, Germany}

\author{Arya Dhar}
\affiliation{Institut f\"ur Theoretische Physik, Goethe-Universit\"at, 60438 Frankfurt am Main, Germany}
\affiliation{Institut f\"ur Theoretische Physik, Leibniz Universit\"at Hannover, Appelstra\ss e 2, 30167 Hannover, Germany}

\author{Tao Qin}
\affiliation{Institut f\"ur Theoretische Physik, Goethe-Universit\"at, 60438 Frankfurt am Main, Germany}

\author{Andreas Gei{\ss}ler}
\affiliation{Institut f\"ur Theoretische Physik, Goethe-Universit\"at, 60438 Frankfurt am Main, Germany}
\affiliation{ISIS, University of Strasbourg and CNRS, 67000 Strasbourg, France}

\author{Mohsen Hafez-Torbati}
\affiliation{Institut f\"ur Theoretische Physik, Goethe-Universit\"at, 60438 Frankfurt am Main, Germany}

\author{Max E.~Sorantin}
\affiliation{Institute of Theoretical and Computational Physics, Graz University of Technology, 8010 Graz, Austria}

\author{Irakli Titvinidze}
\affiliation{Institute of Theoretical and Computational Physics, Graz University of Technology, 8010 Graz, Austria}

\author{Walter Hofstetter}
\affiliation{Institut f\"ur Theoretische Physik, Goethe-Universit\"at, 60438 Frankfurt am Main, Germany}

\date{\today}

\begin{abstract}
In this work we investigate the effect of local dissipation on the presence of density-wave ordering in spinful fermions with both local and nearest-neighbor interactions as described by the extended Hubbard model.
We find density-wave order to be robust against decoherence effects up to a critical point where the system becomes homogeneous with no spatial ordering. Our results will be relevant for future cold-atom experiments using fermions with non-local interactions arising from the dressing by highly-excited Rydberg states, which have finite lifetimes due to spontaneous emission processes.
\end{abstract}


\maketitle

\section{Introduction}
\label{sec:intro}

Coupling to the environment is expected to change the properties of a quantum system. In experiments it usually leads to linewidth broadening~\cite{goldschmidt2016,aman2016}, decoherence and finite lifetime of states~\cite{zeiher2015,zeiher2016}. These dissipative effects are usually limiting experiments. However, coupling between a system and the environment can also lead to exciting new phenomena, such as the quantum Zeno effect~\cite{vidanovic2014,bernier2014,sarkar2014}. Dissipation might also be seen as a tool, allowing to drive the system towards a desired state. Recent proposals include engineering dissipative dynamics to create entangled states~\cite{kraus2008} or to drive the system to Bose-Einstein condensation~\cite{diehl2008}. The study of open quantum systems is also useful for investigating transport properties of quantum dots~\cite{dzhioev2011,dorda2014,schwarz2016,fugger2018} or correlated structures~\cite{ajisaka2012,knap2013,titvinidze2015,titvinidze2016}.

An approach often followed in investigations of open quantum systems involves using well-established methods in quantum optics, e.g.,~the master equation~\cite{breuer2002}\cite{carmichael2002}, to describe dissipation in lattice models, e.g., the Hubbard model~\cite{hubbard1963,gutzwiller1963,konamori1963}, or the Bose-Hubbard model~\cite{gersch1963}. These can be experimentally realized with optical lattices~\cite{jaksch1998,greiner2002,joerdens2008,schneider2008}, which allow for a close comparison between theory and experiment.

The fermionic Hubbard model was originally proposed to study magnetic properties of materials with strong electronic correlations~\cite{varney2009}. Its extended version, including non-local interactions, has been extensively studied due to its relevance for understanding strongly-correlated electronic materials~\cite{micnas1988,dagotto1994,chattopadhyay1997,aichhorn2004,kapcia2017,terletska2017,terletska2018, mckenzie2001,merino2001,kobayashi2004,calandra2002}. In these theoretical investigations phases such as spin density-wave (SDW), charge density-wave (CDW) and charge ordered metals (COM) were observed. Some evidence suggests also that a ``half-metallic'' phase could be found~\cite{garg2014}.

In experiment, long-range interactions between ultracold atoms can be realized in several ways by using dipolar quantum gases~\cite{baranov2012}. Recently, the extended Hubbard model has been realized in experiments with polar molecules~\cite{yan2013} and magnetic atoms~\cite{baier2016}. Another promising approach is one, in which Rydberg excitations~\cite{gallagher1988} are used. Due to the extreme properties of atoms excited to Rydberg state the van der Waals interaction between them can become the dominant energy scale in the system. Loading Rydberg atoms into deep optical lattices has been achieved recently~\cite{schauss2012,schauss2015,zeiher2016,zeiher2017,schauss2018}, allowing the realization of spin-lattice models and observation of spatial ordering due to long-range interaction. Corresponding theoretical investigations for Rydberg atoms in optical lattices have been recently performed predicting crystallization in the frozen limit~\cite{pohl2010,schachenmayer2010,weimer2010,vermersch2015}. Beyond the frozen limit, melting of crystalline structure and formation of supersolid due to kinetic energy has been observed~\cite{lauer2012,geissler2017,li2018}. Effects of dissipation have also been investigated in the frozen limit with a variational principle~\cite{weimer2015}.

However, to our knowledge there has been so far no thorough investigation of the competition between all of the relevant energy scales set by (i) local interaction between atoms, (ii) kinetic energy due to their itinerant nature, (iii) non-local interaction, and (iv) dissipation. The last process is particularly relevant both for experimental realizations of the extended Hubbard model with Rydberg atoms, which are inherently dissipative, and for a better understanding of the possible ordered phases which can appear in open quantum many-body systems.

To investigate this problem we employ the recently developed Lindblad dynamical mean-field theory (L-DMFT)~\cite{knap2013,titvinidze2015,titvinidze2016}. Although DMFT has some limitations due to its local self-energy -- without non-local extensions it cannot describe, e.g.,~\textit{d}-wave superconductivity~\cite{lichtenstein2000} -- it has proven highly successful in the study of correlated lattice problems~\cite{georges1996}. Several approaches have been previously proposed to extend the method to the non-equilibrium regime~\cite{aron2013,aoki2014,li2015}. However, none of these allowed to introduce dissipation on the level of the master equation. 
The L-DMFT method, on the other hand, treats out-of-equilibrium lattice problems by using an appropriately chosen type of impurity solver for the corresponding Anderson impurity model, called the auxiliary master equation approach (AMEA)~\cite{knap2013,dorda2014}.
In contrast to the previous works using this method, which considered a closed quantum system of infinite size, we study a model of an open quantum system. We use the L-DMFT method to investigate the effect of physical dissipation processes on strongly-correlated many-body phases.

The outline of the article is as follows. In Sec.~\ref{sec:ham} we describe in more details the problem which we investigate. In Sec.~\ref{sec:model} we introduce our model Hamiltonian, the extended Hubbard model.
Possible experimental realization of this model is then discussed in Sec.~\ref{sec:rydberg}. The experimentally-relevant dissipative processes that are included in our calculations are introduced in Sec.~\ref{sec:dis}.
A short overview of the L-DMFT technique and its adaptation to dissipative systems is given in Sec.~\ref{sec:ldmft}. We present our results in Sec.~\ref{sec:results} on the competition between density-wave ordering and decoherence. Finally, in Sec.~\ref{sec:end}, results are summarized.

\section{Description of the system}
\label{sec:ham}
\subsection{Model Hamiltonian}\label{sec:model}
\begin{figure}[pt!]
\begin{center}
\resizebox{1.0\columnwidth}{!}{
\includegraphics{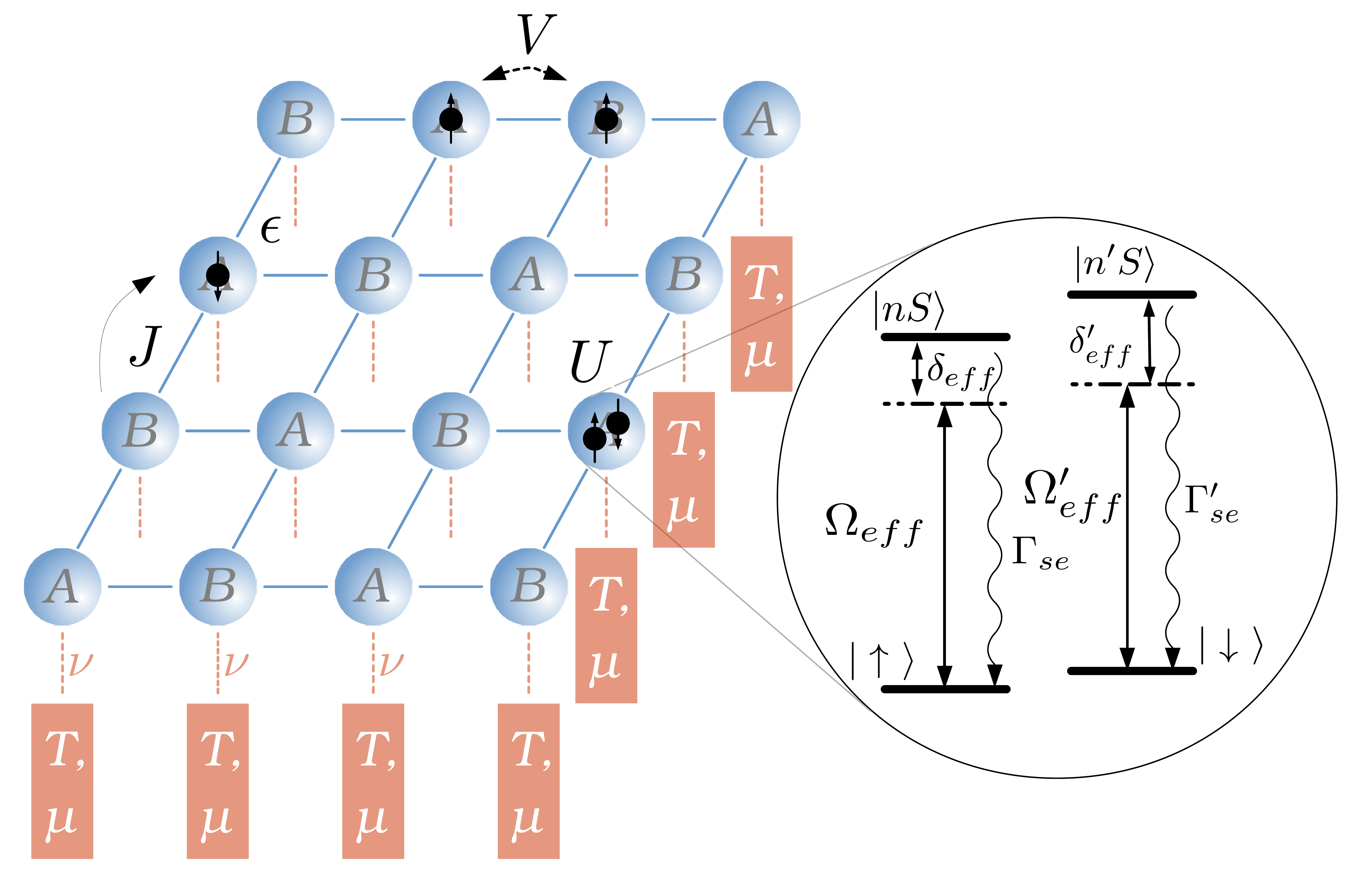}}
\caption{\label{fig:model} Schematic picture of the extended Hubbard model given by the Hamiltonian in Eq.~\eqref{eq:ehm}, including on-site energy, hopping processes, local and non-local interactions, $\epsilon$, $J$, $V$ and $U$ respectively. Each site of the lattice is additionally coupled to an external thermal reservoir, which is described by parameters $T$, $\mu$ and $\nu$, see Eq.~\eqref{eq:fl_dis_rel} and~\eqref{eq:cpl2res} for reference. $A$ and $B$ label the sublattices of this bipartite system. Inset: at each site there is coupling between internal atomic degrees of freedom, which are responsible for Rydberg dressing, as described in App.~\ref{appsec:dress}.}
\end{center}
\end{figure}%

We describe the coherent part of the dynamics by the 2-dimensional (2D) spin-$\frac{1}{2}$ extended Hubbard model (EHM), as illustrated in Fig.~\ref{fig:model}, with the following Hamiltonian
\begin{equation}
\label{eq:ehm}
\begin{split}
\hat{H} & = - J\sum_{\langle i,j\rangle, \sigma} \hat{c}^\dag_{i,\sigma} \hat{c}^{\phantom\dag}_{j,\sigma}
+ \epsilon\sum_{i,\sigma} \hat{c}^\dag_{i,\sigma} \hat{c}^{\phantom\dag}_{i,\sigma} \\
& + U\sum_{i} \hat{n}_{i,\downarrow} \hat{n}_{i,\uparrow}
 + \frac{V}{2}\sum_{\langle i,j\rangle,\sigma,\sigma'} \hat{n}_{i,\sigma} \hat{n}_{j,\sigma'}
 \end{split}
\end{equation}
where index $i$ runs over all lattice sites, $\langle i, j\rangle$ indicates a sum over nearest-neighbor (NN) sites independently, $\hat{c}^{\phantom\dag}_{i,\sigma}$ ($\hat{c}^\dag_{i,\sigma}$) is the fermionic annihilation (creation) operator for a particle on site $i$ with spin $\sigma\in\{\uparrow,\downarrow\}$, $\epsilon$ is the on-site energy, $U$ is the local interaction strength, $V$ is the NN interaction strength, $J$ is the hopping amplitude between NN and $\hat{n}_{i,\sigma}=\hat{c}^\dag_{i,\sigma}\hat{c}^{\phantom\dag}_{i,\sigma}$. In the following we set $\hbar=1$ and $J=1$ as the unit of energy, unless stated otherwise.
At zero temperature and with a Fermi energy $\epsilon_F=0$, the on-site energy $\epsilon$ determines the filling of the system. For example, if it is equal to $\epsilon=-U / 2 - 4 V$ then the system is at half-filling, i.e., at total density $\sum_i \langle \hat{n}_i \rangle= \sum_i ( \langle \hat{n}_{i,\uparrow}\rangle + \langle \hat{n}_{i,\downarrow}\rangle) = N$, with $N$ being the number of sites in the lattice.

The 2D EHM for spin-$\frac{1}{2}$ fermions has been extensively studied in condensed-matter physics, due to its relevance for understanding $d$-wave pairing
and the spatial charge ordering in a wealth of materials. On the square lattice at half-filling, seminal mean-field calculations~\cite{micnas1988,dagotto1994,chattopadhyay1997} showed that there exist only two stable equilibrium phases for the EHM with repulsive interactions, namely the spin density wave (SDW) and charge density wave (CDW) phases, separated by a phase boundary at $V_c = U/4$. More recent beyond-mean-field calculations were also performed with the variational cluster approach~\cite{aichhorn2004}, single-site DMFT~\cite{kapcia2017}, cluster DMFT~\cite{terletska2017}, and GW+EDMFT~\cite{ayral2017}. 
These studies found metallic, Mott-insulating, and charge-ordered phases at half-filling
, but long-range antiferromagnetic order was suppressed at the outset.

At quarter-filling $\sum_i \langle \hat{n}_i \rangle=0.5N$, for both non-zero local and nearest-neighbor repulsion, previous studies have shown that a checkerboard CDW phase appears at large $V$~\cite{mckenzie2001} and a $d_{xy}$-wave superconducting phase at intermediate values of $V$~\cite{merino2001,kobayashi2004}.
In the limit $U,V\gg t$, the ground state was found to be insulating with checkerboard CDW ordering and long-range antiferromagnetism~\cite{mckenzie2001,calandra2002}, based on slave boson calculations and exact diagonalization.

The extended Hubbard model has also been recently investigated beyond half- and quarter-fillings~\cite{kapcia2017,terletska2018}. These works reported the observation of charge-ordered insulator (COI), charge-ordered metal (COM) and Fermi liquid (FL) phases. It was shown that the phase transition from FL phase to quarter-filled COI is discontinuous while the transitions from quarter-filled COI to COM phase and from COM to FL phase are continuous. However, in these studies only non-magnetically ordered phases were considered. 

In the following we work mainly away from half-filling, and focus our study on the above-mentioned charge-ordered phase and its robustness with respect to dephasing processes. 

\subsection{Relevance to experiments with Rydberg-dressed fermions in optical lattices}
\label{sec:rydberg}

The fermionic Hubbard model with only local interactions can be experimentally simulated by ultracold alkali atoms, e.g., $^{40}$K potassium atoms, loaded into optical lattices~\cite{joerdens2008,schneider2008}. 
Such an experimental set-up is indeed very flexible, and allows for the fine-tuning of physical parameters and realization of lattices with different geometries. This allows to explore different parts of the phase diagram of the Hubbard model, including the observation of the Mott insulator and metallic phases~\cite{joerdens2008,schneider2008}.

A non-local interaction of the type present in the EHM can be achieved by coupling fermionic atoms in their ground state to highly-excited Rydberg states~\cite{gallagher1988}. These Rydberg states have a large principal quantum number $n$ which results in exaggerated properties, such as a large radius of the valence electron orbital~\cite{gallagher1988} which scales as $n^2$ and can reach distances on the order of $\sim \mu m$. The large spatial extension of the Rydberg atom, in extreme cases comparable to the typical spacing between lattice sites in an optical lattice, results in significant non-local inter-atom interactions with the van der Waals profile~\cite{singer2005,saffman2010}
\begin{equation}
\label{eq:VvdW}
V_{vdW;ij} = \frac{C_6}{a^6|\textbf{i}-\textbf{j}|^6}
\end{equation}
where $a$ is the lattice spacing, $|\textbf{i}-\textbf{j}|$ the distance between lattice sites $i$ and $j$, and $C_6$ determines the strength of the interaction.
This coupling to a Rydberg state can be realized through interaction of atoms with coherent light of appropriately-chosen frequency and intensity. With each atom represented as a two-level system, the coupling is described in the rotating wave approximation~\cite{cohen1998} by an effective Rabi frequency $\Omega_\mathrm{eff}$ and detuning $\delta_\mathrm{eff}$.
In order to simulate experimentally the EHM for spin-$\frac{1}{2}$, one could focus on the so-called dressing regime, where the effective detuning is much larger than the Rabi frequency $\delta_\mathrm{eff}\gg \Omega_\mathrm{eff}$~\cite{balewski2014}. In this regime a new eigenstate emerges, with properties arising dominantly from the ground state of the atom with a small admixture of properties from the excited Rydberg state. The strength and shape of the interaction potential can then be tuned through the amount of admixture between the two states. The effective potential in this Rydberg-dressing regime is then~\cite{henkel2010}
\begin{equation}
\label{eq:VvdWef}
V_{\mathrm{eff};ij} = \frac{\tilde{C}_6}{a^6|\textbf{i}-\textbf{j}|^6+R_c^6},
\end{equation}
with effective strength $\tilde{C}_6$ and soft-core radius $R_c$. With appropriate choice of experimental parameters, one can reach a regime in which only nearest-neighbor and local interaction processes are relevant for the dynamics. More detailed discussion of the dressed regime can be found in Appendix \ref{appsec:dress}. In particular, the EHM can be realized with $^{40}$K potassium atoms in the dressing regime, with spin obtained by using two different hyperfine states which in turn are coupled with large detuning to the excited Rydberg states (see Fig.~\ref{fig:model}) to realize non-local interaction.

Recently, several experiments with bosonic Rydberg atoms loaded into optical lattice have been performed~\cite{viteau2011,zeiher2016,schauss2012,schauss2015,zeiher2017,schauss2018}. In order to reach longer time scales in the experiments the regime of vanishingly small hopping between lattice sites has been used~\cite{zeiher2016,schauss2012,schauss2015,zeiher2017,schauss2018}. This allowed observation of an emerging ordering in the lattice due to the long-range interaction~\cite{schauss2012,schauss2015,schauss2018}. However, even in the frozen limit, where the Hamiltonian of Rydberg-dressed atoms can be rewritten as an Ising quantum spin model~\cite{schauss2018}, dissipation was already seen as a major obstacle causing, e.g., avalanche loss of particles from the system~\cite{zeiher2016}. In the following, we will go beyond the frozen limit and investigate possible steady-state phases of itinerant atoms which emerge from the full competition between kinetic processes and both short- and long-range interaction.

\subsection{Dissipative processes}
\label{sec:dis}
\subsubsection{Model}
In the above discussion we have focused on \emph{coherent} processes that are present in experiments. However, Rydberg excited states have a relatively short lifetime due to spontaneous emission and black-body radiation~\cite{loew2012,desalvo2016,goldschmidt2016}. 
To take these into account one must include the coupling of the system to its environment, and treat it as a many-body open quantum system. In this paper, we aim in particular at studying the effects of dephasing that occur due to the Rydberg-dressing.

As is often the case in the theory of open quantum systems, we will use the Born-Markov approximation to describe the evolution of the system. 
As a result, we can use the Lindblad master equation~\cite{breuer2002,carmichael2002}
\begin{equation}\label{eq:liouvillian}
\frac{d \hat{\rho}}{d t} = -i\left[\hat{H}, \hat{\rho} \right] + \doublehat{\mathcal{L}}[\hat{\rho}]
\end{equation}
where $\hat{\rho}$ is the density matrix operator, and $\doublehat{\mathcal{L}}$ is the superoperator which describes dissipation. In the Lindblad equation it is defined according to
\begin{equation}\label{eq:liouvillian2}
\doublehat{\mathcal{L}}[\hat{\rho}] = \frac{1}{2} \sum_{\mu\nu} \Gamma_{\mu,\nu} \left( 2 \hat{L}_\nu \hat{\rho} \hat{L}^\dagger_\mu - \left\{ \hat{L}^\dagger_\mu \hat{L}_\nu, \hat{\rho} \right\} \right),
\end{equation}
where $\hat{L}_\mu$ are jump operators, $\Gamma_{\mu\nu}$ are dissipation coefficients and $\mu$, $\nu$ iterate over relevant quantum numbers.
Regarding the relevant jump operators to include in our description, we assume here that the dominant dissipative effects for Rydberg atoms are spontaneous emission processes with rate $\Gamma_\mathrm{se}$ (see Fig.~\ref{fig:model}). This can be mapped within the dressing regime to a dephasing process and described by the effective dephasing rate $\Gamma_\mathrm{dp}$ and the following jump operator (see App. \ref{appsec:dress})
\begin{equation}
\label{eq:Ldp}
\hat{L}_{{dp},i,\sigma} = \hat{c}_{i,\sigma}^\dag\hat{c}^{\phantom\dag}_{i,\sigma}.
\end{equation}
Note that with this type of jump operator the time evolution given by Eq.~\eqref{eq:liouvillian} is effectively quartic in terms of creation and annihilation operators. Therefore, effects of dissipation are incorporated into the self-energy, together with the effects of interaction.

Such dephasing terms conserve the local particle number and hence cannot change the local occupation.
One of the effects of this type of dissipation on the many-body state is, however, to cause the decay of the off-diagonal elements of the density matrix and drive the system towards the infinite-temperature state in the absence of an external thermal bath (cf.~Appendix~\ref{appsec:disheat} and \cite{sarkar2014,bernier2014}).

Dephasing terms of the form in Eq.~\eqref{eq:Ldp} correspond, in the theory of open quantum systems, to a continuous measurement process of the site occupation variable~(see \cite{breuer2002}, Ch.~3.5).
Such terms are also useful to describe the coupling of the local fermion density to the environment through non-local interactions~\cite{sarkar2014,bernier2014}.

\section{Non-Equilibrium DMFT: auxiliary master equation approach}
\label{sec:ldmft}
The auxiliary master equation approach (AMEA) to non-equilibrium DMFT, here referred to as L-DMFT, has been proposed and developed to study transport properties of a correlated electronic layer coupled to non-interacting leads~\cite{knap2013,titvinidze2015,titvinidze2016}. However, we show here that the method has potential for applications to other types of out-of-equilibrium problems, e.g.,~a lattice system with local dissipation. We describe how to adapt the method to such a problem and apply it to the extended Hubbard model with local dephasing. As described in the preceding section, this model can be simulated with Rydberg atoms loaded into optical lattices, where the dissipative processes are naturally present.

The L-DMFT method allows to find the steady-state of a system far-from-equilibrium in a self-consistent way and then to calculate static and dynamic local quantities. Due to the self-consistent approach it is better to have a unique steady-state. If we were to consider a closed system described by the Hamiltonian of Eq.~\eqref{eq:ehm}, we would face the problem of non-unique steady-states. Indeed, for a closed system there are as many steady-states as there are eigenstates of the Hamiltonian. As we are interested in an open quantum system, where the lattice is subject to dephasing, this issue should not occur. While the dissipation indeed renders the steady-state unique, it also generically heats the system and drives it towards an uninteresting, infinite-temperature steady-state~\cite{bernier2014} (App. \ref{appsec:disheat}).

In order to have a non-trivial steady-state we assume that each site of the extended Hubbard model is coupled to a separate heat and particle reservoir that is always in thermal equilibrium, as depicted in Fig.~\ref{fig:model}. The local reservoirs act as a heat drain and allow us to study the limit of vanishing dissipation strength by lifting the degeneracy of the steady-states.
We will use an exactly solvable model for the bath known as Davies' model of heat conduction~\cite{davies1978} or B\"uttiker's heat-bath model~\cite{buttiker1985,buttiker1986}.
This heat-bath model was recently employed to study the effect of dissipation on interacting many-body systems, using a variant of non-equilibrium DMFT~\cite{tsuji2010,aron2013,aoki2014,li2015,qin2018} and quantum Monte Carlo path integrals~\cite{yan2018}. To put the additional thermal baths into context we note that they are commonly used in context of the Floquet-DMFT to achieve \textit{inhomogeneous} equations with a unique, non-trivial solution of the equations for the steady-state~\cite{aoki2014,qin2018}. Without heat-baths a periodically driven system approaches an infinite temperature state in the long timescale, but in the intermediate timescale it can be found in a quasistationary, Floquet prethermalized state~\cite{peronaci2018}. Overall, heat-bath can be treated there as a theoretical ``trick'' for numerical methods, which allows to study the intermediate timescale state as a steady-state. We also note that recently an experiment was performed, in which an optical lattice was coupled to a thermal reservoir of atoms captured in a magneto-optical trap~\cite{chong2018}.

The local thermal baths are assumed to be one-dimensional semi-infinite chains of non-interacting fermions with hopping $J_b$ between neighboring sites and a retarded Green function given by
\begin{equation}\label{eq:leadGf}
g_b^R(\omega) = \frac{\omega}{2 J_b^2}-\rmi\frac{\sqrt{4J_b^2-\omega^2}}{2J_b^2}.
\end{equation}
As the bath is assumed to be always in thermal equilibrium, its Keldysh Green function is given by the fluctuation-dissipation theorem
\begin{equation}\label{eq:fl_dis_rel}
g_b^K(\omega)  = 2\rmi\left[ 1 -2f_b(\omega) \right]\mathrm{Im}\{g_b^R(\omega)\},
\end{equation}
where $f_b(\omega)$ is the Fermi-Dirac distribution. We set the Fermi energy (chemical potential) $\epsilon_F=0$ and the temperature $T=0$. The value of hopping in the thermal baths is set to $J_b=7.5$ which gives a half-bandwidth ($2J_b$) on the order of magnitude of the maximal considered value of $U$. This allows for thermalization in a broad energy spectrum even in the presence of the Hubbard band splitting due to the local interaction. The coupling of the local thermal baths to the system is realized via exchange of particles with hopping amplitude $\nu$. The particular form of this coupling will be introduced in the next section.


\subsection{DMFT self-consistency}
\label{subsec:dmftsc}

In DMFT a single approximation is made that the self-energy is a purely local quantity~\cite{georges1996}, such that
\begin{equation}
\Sigma_{ij,\sigma}(\omega) = \delta_{ij} \Sigma_{i,\sigma}(\omega),
\end{equation}
where $i$ and $j$ are lattice indices. As a consequence one is able to map a full lattice problem onto a set of local effective quantum impurity models, which significantly reduces the size of the many-body problem while fully preserving the nature of local quantum correlations. These impurity problems are coupled in the self-consistent approach via a Dyson equation given further in text.

Due to the local character of DMFT, we must however treat the non-local nearest-neighbor interaction term in Eq.~\eqref{eq:ehm} within a Hartree mean-field approximation, where the interaction operator is mapped onto
\begin{equation}
\frac{V}{2}\sum_{\langle i,j\rangle,\sigma} \hat{n}_{i,\sigma} \hat{n}_{j,\sigma} \to
V\sum_{\langle i,j\rangle,\sigma} \hat{n}_{i,\sigma} \langle \hat{n}_{j,\sigma} \rangle
\end{equation}
with $\langle \hat{n}_{j,\sigma}\rangle$ determined self-consistently.

The model which we consider is translationally invariant, which allows to find a symmetry between lattice sites and reduce the size to a small number of inequivalent impurity problems. However, the symmetry of the ground state on the lattice can be spontaneously reduced in certain ordered phases such as the CDW phase, which we want to investigate. We assume throughout this work that the system has two translationally invariant sublattices, $A$ and $B$, which results in two different impurity problems to solve. The sublattices are defined such that each site from sublattice $A$ is neighboring only with sites belonging to sublattice $B$ and vice versa, see Fig.~\ref{fig:model}.

The derivation of the DMFT equations in the non-equilibrium Keldysh formalism is well established in the literature~\cite{knap2013,aoki2014,titvinidze2015}. We refer the reader in particular to Ref.~\onlinecite{titvinidze2015}, while here we present the differences with respect to this reference, which arise from the two-sublattice structure and geometry of the system that we consider here.

We use a notation for the Green function in which
\begin{equation}
\mathbf{G} = \begin{pmatrix} G_{AA} & G_{AB} \\ G_{BA} & G_{BB}\end{pmatrix},
\end{equation}
where the $A$ and $B$ indices mark each sublattice. On top of that, to introduce retarded ($R$), advanced ($A$) and Keldysh ($K$) components of the Green functions, we use the notation
\begin{equation}
\udln{\mathbf{G}} = \begin{pmatrix} \mathbf{G}^R & \mathbf{G}^K \\ \mathbf{0} & \mathbf{G}^A \end{pmatrix}.
\end{equation}
Due to the translational invariance of the sublattices $A$ and $B$ one can perform a Fourier transform with a reduced Brillouin zone ($BZ'$).
In analogy to Ref.~\cite{titvinidze2015} we express the Green functions in momentum space. The Green functions of the non-interacting model decoupled from the thermal bath is given by
\begin{equation}
\mathbf{g}_0^R(\vec{k},\omega) = \begin{pmatrix} \omega + \rmi 0^+ -\epsilon & -E_c(\vec{k}) \\ -E_c(\vec{k}) & \omega + \rmi 0^+ - \epsilon
\end{pmatrix}^{-1},
\end{equation}
where $E_c(\vec{k})=-2J(\cos(k_x)+\cos(k_y))$, and with wave-vectors $\vec{k}$ from the reduced Brillouin zone. Correspondingly, as the local baths are decoupled from one another, their Green functions after Fourier transform have the form
\begin{equation}\label{eq:ABnot}
\begin{split}
\mathbf{g}_b^{R/K/A} (\vec{k},\omega) & = \mathbf{g}_b^{R/K/A} (\omega)\\
& = \begin{pmatrix}
g_b^{R/K/A}(\omega) & 0 \\ 0 & g_b^{R/K/A}(\omega)
\end{pmatrix}.
\end{split}
\end{equation}
We note that the Keldysh part $\mathbf{g}^K_0(\vec{k},\omega)$ is state dependent and not uniquely defined because it corresponds to a system decoupled from any thermal bath. Nevertheless, we will only need to use the inverse of this Green function, for which the Keldysh part $\left[\mathbf{g}^{-1}_0(\vec{k},
\omega)\right]^K$ is infinitesimally small~\cite{knap2013}\footnote{If we neglect the Keldysh part in the decoupled system it is essential that the full system can thermalize either through interaction or through coupling to thermal bath.}.

Based on the above Green functions, we can determine the Green functions of a two-dimensional non-interacting system coupled to the thermal bath. We get
\begin{equation}\label{eq:cpl2res}
\udln{\mathbf{G}}_0^{-1} (\vec{k},\omega) = \udln{\mathbf{g}}_0^{-1}(\vec{k},\omega) - \nu^2 \udln{\mathbf{g}}_b (\omega).
\end{equation}
This equation determines the form of the coupling between the lattice sites and the thermal reservoir with $\nu^2$ defining the coupling strength.
The Green function of the interacting model coupled to thermal baths is then given by the Dyson equation~\cite{georges1996,knap2013,titvinidze2015}
\begin{equation}
\udln{\mathbf{G}}^{-1} (\vec{k},\omega) = \udln{\mathbf{G}}_0^{-1}(\vec{k},\omega) - \udln{\mathbf{\Sigma}} (\omega).
\end{equation}
Here $\udln{\mathbf{G}}^{-1} (\vec{k},\omega)$ is the Green function of the full system in momentum space. The self-energy $\udln{\mathbf{\Sigma}} (\omega)$ here describes effects of local and non-local interaction (the latter on the mean-field level) as well as the effects of dephasing. As it is assumed to be local, it is also momentum independent in the reduced Brillouin zone, but it might be different for sublattices $A$ and $B$
\begin{equation}
\mathbf{\Sigma}^{R/K/A}(\omega) = \begin{pmatrix} \Sigma^{R/K/A}_A(\omega) & 0 \\ 0 & \Sigma^{R/K/A}_B(\omega).
\end{pmatrix}
\end{equation}
To close the self-consistency equations we extract the local part of the lattice Green function, i.e., for the sublattice $A$ ($B$) we have
\begin{equation}\label{eq:loc_Green_f}
\udln{G}_{A(B)}(\omega) = \int_{BZ'} \frac{\d \vec{k}}{(2\pi)^2}\udln{G}_{A(B)}(\vec{k},\omega).
\end{equation}
Note that while here we perform the operation for sublattice $A$ and $B$ separately, we still have a matrix equation with retarded, advanced and Keldysh parts. We use the above result in the local Dyson equation, which reads
\begin{equation}\label{eq:locDyson}
\udln{\Delta}_{A(B)}(\omega) = \udln{G}_{0,A(B)}^{-1}(\omega) - \udln{\Sigma}_{A(B)}(\omega) - \udln{G}^{-1}_{A(B)}(\omega),
\end{equation}
with $\udln{\Delta}_{A(B)}$ being the hybridization function which describes the effect of coupling the non-interacting impurity to both the thermal bath and the interacting lattice that surrounds it, and with
\begin{equation}
G^R_{0,A}(\omega) = G^R_{0,B}(\omega) = \left(\omega + \rmi 0^+ - \epsilon\right)^{-1} .
\end{equation}
representing the local Green function of the single, non-interacting lattice site decoupled from both the thermal bath and the surrounding lattice. The Keldysh part of the inverse Green function is again negligible.

The remaining problem is to solve the two emerging impurity problems for sublattice $A$ and $B$, given the hybridization functions $\udln{\Delta}_{A(B)}$. Once one can do this, one can solve the full problem self-consistently.

\subsection{Impurity solver: auxiliary master equation approach}
\label{subsec:amea}

Solving the impurity problems is usually the bottleneck of the DMFT method. Here we deal with two independent problems, one for each of the sublattices. In the following the sublattice index $\alpha\in\{A,B\}$ denotes which impurity problem we consider. What significantly adds to the complexity of the task is the fact that the system is not in thermal equilibrium (at least in the general case) but rather in a steady-state of some non-trivial dissipative dynamics.

A method well-suited to solve such an impurity problem is the auxiliary master equation approach (AMEA)~\cite{knap2013,dorda2014,titvinidze2015}. In our implementation of non-equilibrium DMFT, we adapt it to a problem with physical local dissipation. Below we briefly list the main points of this method, focusing on what is most relevant to our problem.

\begin{figure}[pt!]
\begin{center}
\resizebox{0.9\columnwidth}{!}{
\includegraphics{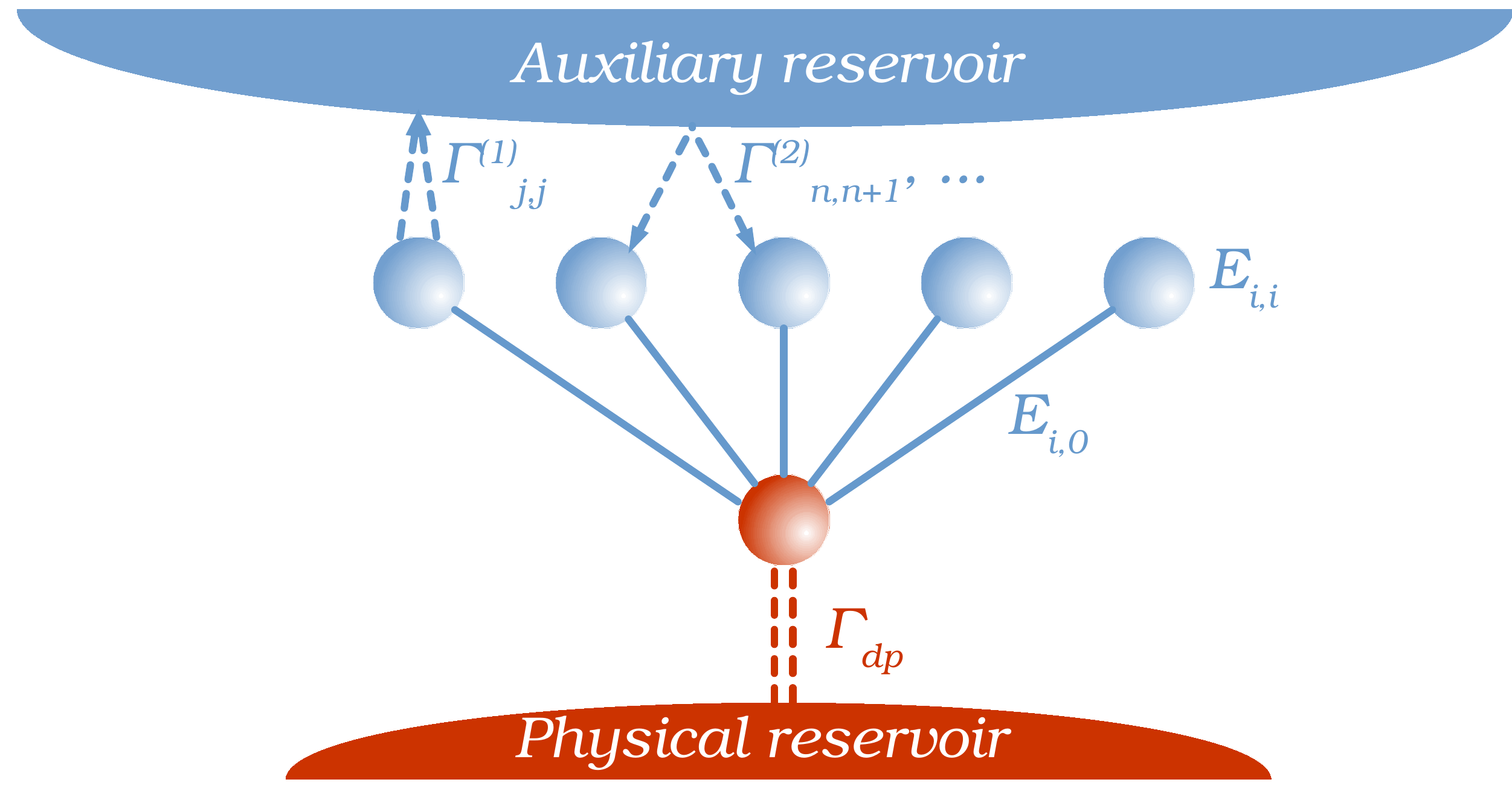}}
\caption{\label{fig:amea} Auxiliary system for an impurity embedded in lattice. Index $i=0$ corresponds to the impurity and indices $i>0$ correspond to auxiliary bath sites. Parameters $E_{ij}$ represent the on-site energies and the hopping amplitudes in the impurity problem, see Eq.~\eqref{eq:aux_ham}. Parameters $\Gamma_{ij}^{(1/2)}$ describe dissipative coupling between auxiliary bath sites and auxiliary reservoir, see Eq.~\eqref{eq:aux_dis}, while $\Gamma_\mathrm{dp}$ describes physical dissipation.}
\end{center}
\end{figure}

The foundation of this method is laid by the exact diagonalization approach to the impurity problem~\cite{georges1996}, in which it is mapped onto an effective finite size problem. A single impurity is coupled to a finite number $N_b$ of non-interacting bath sites, which imitate the surrounding of the impurity as closely as possible. These bath sites are completely auxiliary and should not be confused with the thermal bath introduced at the beginning of this section. The effective impurity Hamiltonian reads
\begin{equation}\label{eq:aux_ham}
H_{\alpha,aux} = \sum_{i,j=0;\sigma}^{N_b} E_{\alpha,ij} \hat{d}^\dag_{i,\sigma} \hat{d}_{j,\sigma} + U\hat{d}^\dag_{0,\downarrow} \hat{d}^\dag_{0,\uparrow}\hat{d}_{0,\uparrow} \hat{d}_{0,\downarrow},
\end{equation}
where $\hat{d}_{i,\sigma}$ is the annihilation operator on site $i$ with spin $\sigma$, the $i$ and $j$ indices run over all possible sites in auxiliary impurity problem with $i=0$ referring to the impurity. $E_{\alpha,ij}$ are arbitrary parameters subject to the constraint that they should form a hermitian matrix and with the value of $E_{\alpha,00}$ fixed by the original lattice problem~\footnote{It includes effects of non-local interaction treated on the mean-field level}. We also choose to work with a star geometry of the bath, without loss of generality, cf. Fig. \ref{fig:amea}. However, this Hamiltonian is by itself not sufficient to describe the time evolution of an open quantum system.

To circumvent this issue, in the AMEA approach the bath sites are coupled to a Markovian auxiliary reservoir, see Fig. \ref{fig:amea}. This allows to describe the time evolution with the Lindblad master equation Eq.~\eqref{eq:liouvillian} in which the Hamiltonian is given by Eq.~\eqref{eq:aux_ham}. The dissipative part of the master equation is determined by two terms. The first term is the local, physical dissipation given in Eq.~\eqref{eq:liouvillian2} which acts only on the impurity state with index 0. The second term is determined by
\begin{equation}\label{eq:aux_dis}
\begin{split}
\doublehat{\L}_{\alpha,aux}\left[\hat{\rho}\right] & =  \sum_{\sigma;i,j=1}^{N_b} 2 \\
& \times\left[ \Gamma^{(1)}_{\alpha,ij}\left( \hat{d}_{i,\sigma}^{\phantom\dag} \hat{\rho} \hat{d}_{j,\sigma}^\dag - \frac{1}{2}\{\hat{\rho},\hat{d}^\dag_{j,\sigma} \hat{d}_{i,\sigma}^{\phantom\dag} \} \right)\right.\\
&  +\left.\Gamma^{(2)}_{\alpha,ij}\left( \hat{d}^\dag_{j,\sigma} \hat{\rho} \hat{d}_{i,\sigma}^{\phantom\dag} - \frac{1}{2}\{\hat{\rho},\hat{d}_{i,\sigma}^{\phantom\dag} \hat{d}_{j,\sigma}^\dag \} \right) \right]
\end{split}
\end{equation}
and describes the coupling of the bath sites to the Markovian reservoir. Here $\Gamma_{\alpha,ij}^{(1)}$ and $\Gamma_{\alpha,ij}^{(2)}$ are arbitrary parameters subject to the constraint that they should form a hermitian, positive-definite matrix.

This type of non-equilibrium impurity model has been extensively studied in the literature~\cite{dzhioev2011,ajisaka2012,dorda2014,schwarz2016}. Using exact diagonalization of the Liouvillian in the super-fermionic representation (which doubles the Hilbert space limiting achievable $N_b$) one can solve this impurity problem~\cite{knap2013,dzhioev2011super}. Note also that if we switch off the local interaction and dissipation on the impurity, the model becomes quadratic and therefore analytically solvable. Consequently, one can calculate the effective hybridization function $\udln{\Delta}_{aux}$ with little computational effort~\cite{knap2013}.

The free parameters $E_{\alpha,ij}$ and $\Gamma_{\alpha,ij}^{(1/2)}$ of the impurity model form a set of variables $\{x_\alpha\} = \bigcup_{ij}\{E_{\alpha,ij}, \Gamma^{(1)}_{\alpha,ij}, \Gamma^{(2)}_{\alpha,ij}\}$. This set can be further reduced, cf. Ref~\cite{knap2013}, using symmetries of $\Gamma$ matrix and an appropriate geometry of the impurity model, e.g., the star geometry from Fig.~\ref{fig:amea}. The values of the variables are chosen in such a way that the cost function
\begin{equation}\label{eq:cost_fun}
\begin{split}
& \chi_\alpha\left(\{x_\alpha\} \right) \\
& = \int_{-\infty}^\infty \d \omega \left[ \chi_\alpha^R(\omega,\{x_\alpha\}) + \chi_\alpha^K(\omega,\{x_\alpha\})+ \chi_\alpha^f(\omega,\{x_\alpha\}) \right]
\end{split}
\end{equation}
is minimized. Different contributions to $\chi$ are defined as
\begin{equation}\label{eq:cost_fun_el}
\begin{split}
\chi_\alpha^R\left(\omega,\{x_\alpha\} \right) & = \mathrm{Im}\left[\Delta_\alpha^R(\omega) - \Delta^R_{\alpha,aux}\left(\omega,\{x_\alpha\}\right)\right]^2, \\
\chi_\alpha^K\left(\omega,\{x_\alpha\} \right) & = \mathrm{Im}\left[\Delta_\alpha^K(\omega) - \Delta^K_{\alpha,aux}\left(\omega,\{x_\alpha\}\right)\right]^2, \\
\chi_\alpha^f\left(\omega,\{x_\alpha\} \right) & = \left|f_\alpha(\omega) - f_{\alpha,aux}\left(\omega,\{x_\alpha\}\right)\right|^2 |\mathrm{Im}[\Delta^R_{\bar{\alpha}}(\omega)]|.
\end{split}
\end{equation}
The bar in $\bar{\alpha}$ denotes the complement of $\alpha$ in the set $\{A,B\}$, $\Delta_\alpha(\omega)$ is the physical hybridization function for sublattice $\alpha$ obtained from the Dyson equation~\eqref{eq:locDyson}, $\Delta_{\alpha,aux}(\omega,\{x_\alpha\})$ is the auxiliary hybridization function for sublattice $\alpha$ in the impurity model~\cite{knap2013,titvinidze2015}, $f_\alpha(\omega)$ is the distribution function calculated using the fluctuation-dissipation relation and reads
\begin{equation}\label{eq:delta2distr}
f_\alpha(\omega) = \frac{1}{2} - \frac{1}{4}\frac{\mathrm{Im}[\Delta^K_\alpha(\omega)]}{\mathrm{Im}[\Delta^R_\alpha(\omega)]}.
\end{equation}
An analogous formula is used for $f_{\alpha,aux}(\omega)$.

The terms $\chi_\alpha^R$ and $\chi_\alpha^K$ are responsible for obtaining the best fit of the retarded and Keldysh parts of the hybridization function, respectively. However, in the case of a small number of bath sites, the accuracy of the fit for these terms might come at the cost of a less accurate reproduction of the distribution function. This is compensated by our inclusion of the last term, $\chi^f_\alpha$. Obtaining an accurate fit of the distribution function is necessary only in the region of significant spectral weight, hence the factor $|\mathrm{Im}[\Delta_{\bar{\alpha}}(\omega)]|$.


\subsection{Limitations}
\label{subsec:limit}

It is clear that the approximation made in the course of the AMEA gets better with increasing number of bath sites $N_b$. However, due to the exponential scaling of the size of the problem with the number of bath sites and doubling of the Hilbert space one cannot reach large values of $N_b$ when using an exact diagonalization based solver. Within our implementation we are able to set this parameter up to $N_b=5$. In other works a number of bath sites up to $N_b=6$ has been reported~\cite{dorda2014,titvinidze2015,titvinidze2016}. However, higher number of baths sites in the L-DMFT has not been reached yet. An alternative to the exact diagonalization based solver in the AMEA impurity problem is to use the matrix product states approach~\cite{dorda2014,schwarz2016,fugger2018}. With this method values of up to $N_b=20$ have been reached~\cite{fugger2018}. However, currently combining this method and the DMFT self-consistency is not practical, as the computational effort to solve a single impurity problem is too large to be used in a self-consistent approach.

With a limited number of bath sites some physical quantities might not be recovered accurately. For example, upon investigation of the occupation of different energy states, the Fermi distribution in thermal equilibrium might not be reproduced precisely. This leads to deviations between the results obtained with a standard, equilibrium DMFT solver, which can reach higher accuracy, and the L-DMFT solver used here.

One of the features that are difficult to capture using a small number of bath sites is the magnetic response of the system. E.g., with $N_b=5$ we could not reproduce the anti-ferromagnetic phase of the standard Hubbard model at zero temperature. Nevertheless, using the AMEA within stochastic wave function approach~\cite{sorantin2018} we checked that the discrepancy between the results of the equilibrium and AMEA impurity solvers is decreasing with increasing number of bath sites. Also for other types of impurity magnetic response the value of $N_b>10$, which might be reached with the matrix product states method, was enough to get accurate results~\cite{fugger2018}.

Another relevant effect occurs if one of the impurity energy levels lies outside of the band specified by the heat-bath. In such case a localized state appears, whose evolution is not captured within the Markov approximation of the auxiliary reservoir~\cite{zhang2012}. This issue is here amended by the choice of the local thermal baths with a broad energy spectrum.

\section{Results}
\label{sec:results}
\subsection{Equilibrium}
\label{sec:eq}

We begin with the investigation of a system in thermal equilibrium, i.e., without dephasing process. As in this regime one can employ alternative methods to study the model described by Eq.~\eqref{eq:ehm}, this serves both as a reference for the calculation with dephasing and as a benchmark of the L-DMFT method. We will compare it to the equilibrium DMFT and Hartree-Fock methods.
Note that we allow only for two phases: charge density wave and normal phase with homogeneous particle density and neglect any magnetic ordering possibly emerging in the system, such that $\langle\hat{n}_{i,\uparrow}\rangle = \langle \hat{n}_{i,\downarrow}\rangle$.

In our equilibrium DMFT and Hartree-Fock calculations we set the chemical potential to $\mu=0$, temperature to $T=0$ and switch off the coupling to external thermal baths. For the L-DMFT calculations the chemical potential and the temperature are set to the same values indirectly, through coupling to the thermal reservoir, see Eq.~\eqref{eq:fl_dis_rel} and \eqref{eq:cpl2res}. We set the strength of this coupling to $\nu^2=0.5$. To obtain both CDW and normal phases we set the NN-interaction strength to $V=2$. We perform calculations with local-interaction strength ranging from $U=1$ to $U=16$. On-site energy is set to $\epsilon=-\frac{U}{2}$. This would correspond to half-filling in the absence of non-local interaction. However, with $V=2$ the filling is lower and thus results are away from half-filling.

\begin{figure}[pt!]
\begin{center}
\resizebox{1.0\columnwidth}{!}{
\includegraphics{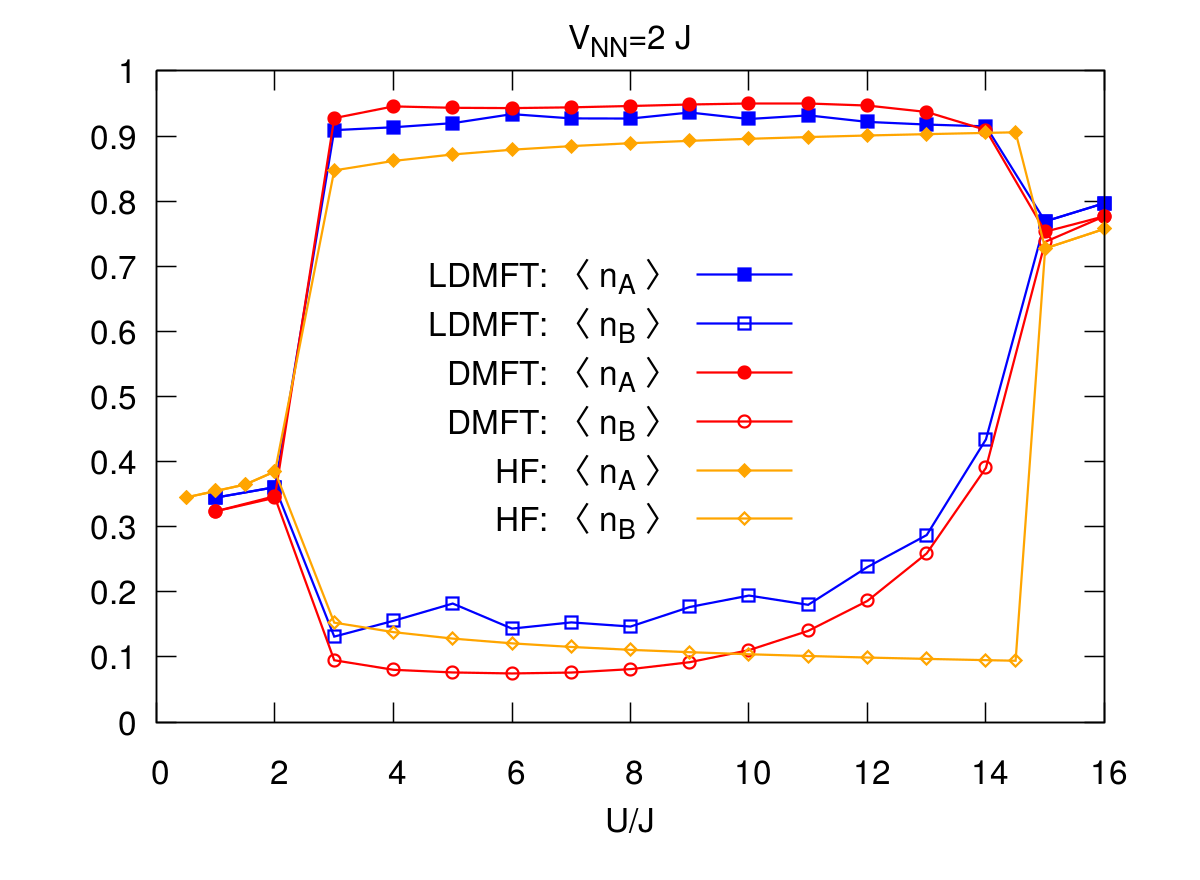}}
\caption{\label{fig:occ_eq} Comparison of local occupations $\langle\hat{n}_{A(B)}\rangle$ obtained within equilibrium Hartree-Fock (HF, orange diamonds), equilibrium DMFT (red circles) and L-DMFT (blue squares). Index $A$ ($B$) corresponds to quantities in sublattice $A$ ($B$). System parameters are set to $J=1$, $\epsilon=-\frac{U}{2}$, $V=2$, $\Gamma_\mathrm{dp}=0$. Local thermal bath parameters in L-DMFT are: coupling strength $\nu^2=0.5$, temperature $T=0$ and chemical potential $\mu=0$. DMFT and HF results are obtained for $T=0$ and $\mu=0$, without coupling to local thermal baths.}
\end{center}
\end{figure}%

In Fig.~\ref{fig:occ_eq} we present the comparison of local occupations for a single spin species, $\langle\hat{n}_{A(B)}\rangle=\langle\hat{n}_{A(B),\downarrow}+\hat{n}_{A(B),\uparrow}\rangle$, obtained within equilibrium DMFT and Hartree-Fock mean-field. The equilibrium DMFT results were obtained with the exact diagonalization impurity solver~\cite{hafez-torbati2018}. We observe that for intermediate values of the interaction strength a checkerboard CDW phase emerges, resulting in spontaneous symmetry breaking with non-zero value of the checkerboard 
order parameter $\Delta n=|\langle \hat{n}_A-\hat{n}_B\rangle|$.

The phase transition at high $U$ (and as a result high $|\epsilon|$) occurs due to the competition between the on-site energy $\epsilon$ and NN interaction $V$. Approximately, the energy cost of adding a particle at (almost empty) sublattice $B$ is given by $4V=8J$ (due to four singly occupied neighbors) and the energy gain is given by $\epsilon=-U/2$. Therefore, in the atomic limit one can expect a phase transition around $U=16J$. The hopping processes lead to hybridization of the two sublattices. This results in decreasing value of the checkerboard order parameter $\Delta n$ as we approach the phase transition point, and a shift of the critical interaction strength $U_c$ to lower values of around $U_c=14.5\pm0.5$.
At low values of $U$ the phase transition occurs around $U_c=2.5\pm0.5$ with a jump in the total filling. As all four energy scales, namely $U$, $J$, $V$ and $\epsilon$, are comparable and we have not been able to find a simple explanation for the nature of this phase transition.

To benchmark the L-DMFT technique we performed a series of tests for an arbitrarily chosen value of on-site interaction $U=8$. Firstly we investigated how the accuracy of the method depends on the number of bath sites $N_b$ used in the impurity solver. We checked how the spectral function and the filling of sublattices changes for $1\leqslant N_b \leqslant 5$ (results not shown in here). While we observe significant deviations for $N_b\leqslant 2$, the values $N_b=3,\ 4$ and $5$ give comparable results.

\begin{figure}[pt!]
\begin{center}
\resizebox{1.0\columnwidth}{!}{
\includegraphics{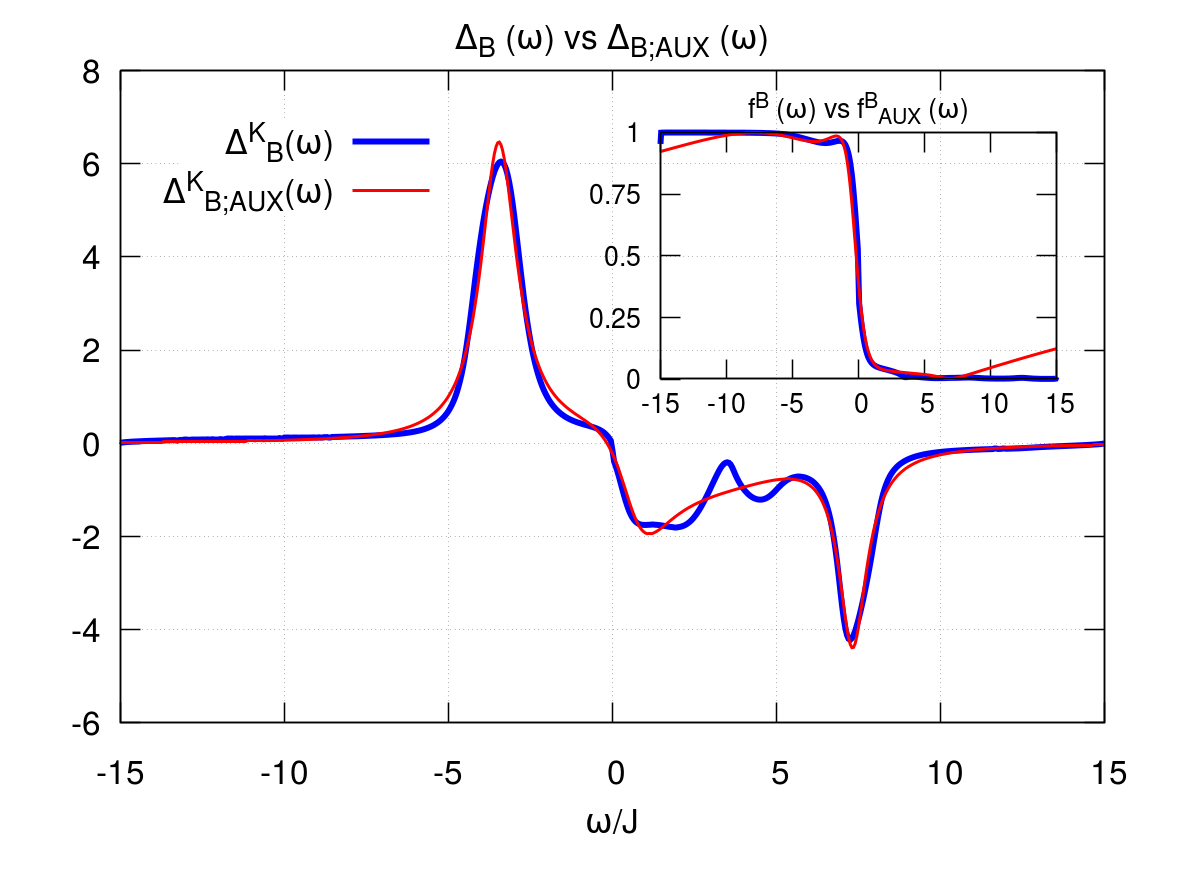}}
\caption{\label{fig:hyb} Comparison of the physical (blue thick line) and auxiliary (red thin line) hybridization functions in the converged solution with $U=8$ (other parameters are the same as in Fig.~\ref{fig:occ_eq}) for sublattice $B$: $\Delta^{K}_B(\omega)$, $\Delta^{K}_{B;aux} (\omega)$. Inset: comparison of the distribution functions $f^B(\omega)$ extracted from the hybridization function according to Eq.~\eqref{eq:delta2distr}.}
\end{center}
\end{figure}%
Next we focused on the performance of the method with $N_b=4$. In Fig. \ref{fig:hyb} we present the comparison of the hybridization functions at $U=8$ for sublattice $B$ obtained from the Dyson equation \eqref{eq:locDyson} and after mapping onto the impurity AMEA model. We observe that while some finer details are lost in the mapping procedure, the main features are properly reproduced. Upon closer investigation of the distribution functions $f_{B,aux}(\omega)$ describing the environment of a site from sublattice $B$ (displayed in Fig.~\ref{fig:hyb}), we notice that: (i) the auxiliary hybridization function does not reproduce perfectly the Fermi-Dirac distribution function, (ii) there are small discrepancies between the physical and auxiliary distribution functions in the region where the spectral weight $\mathrm{Im}[\Delta^R(\omega)]$ is large. Discrepancies in the remaining regions do not lead to significant issues -- as the spectral weight is small in these regions it does not contribute strongly to the dynamics or the total occupation of the system.

The discrepancies in the reproduced distribution function might lead to different values of occupation and double occupancy between the equilibrium and non-equilibrium solvers. This issue is indeed observed when comparing the DMFT and L-DMFT results (cf.~Fig.~\ref{fig:occ_eq}). Nevertheless, it is minimized by an appropriate choice of the cost function, which minimizes the error in the relevant regions through $\chi^f_\alpha$.

The discrepancies in the hybridization functions in the AMEA have yet another consequence. As not all features are perfectly reproduced, there might be local mi\-ni\-ma of the cost function \eqref{eq:cost_fun}, in which features which are captured more accurately appear in different parts of the spectrum. As a result, one might expect more than one self-consistently converged solution of the full L-DMFT approach. We checked that this leads to at most small quantitative differences in the converged solution. Qualitative features of the results remain unchanged.

Having established that the method gives a good qualitative description of the system we compared the results of equilibrium DMFT with those of L-DMFT for a wide range of values of $U$, Fig.~\ref{fig:occ_eq}. 
The non-smoothness of the L-DMFT results originates from the emergence of multiple self-consistent solutions discussed above and from discrepancies in the effective distribution function $f(\omega)$. We observe that the methods yield similar results, but one can observe some quantitative differences. In all cases the system does not exhibit a CDW phase for weak local interaction. As $U$ is increased the system undergoes a phase transition, which occurs at a critical value around $U_c\approx 2.5\pm 0.5$ with a sharp change of the order parameter $\Delta n$. Investigating the type of this phase transition goes beyond the scope of this paper. As the value of $U$ is further increased the order parameter decreases until it vanishes completely at around $U_c\approx 14.5\pm0.5$. Overall, the comparison of DMFT and L-DMFT shows that the latter agrees qualitatively with an equilibrium method which is well established in the literature and which captures effects of strong local correlations. Quantitative differences can serve as a measure of the accuracy for our method.

To check the effect of exchange (Fock) terms due to the nearest-neighbor interaction, which are absent in DMFT, we additionally perform a mean-field study of the ground state long-range order of the Hamiltonian in Eq.~\eqref{eq:ehm}, by following the self-consistent mean-field method that includes both Hartree- and Fock-decoupling of local and non-local interaction terms~\cite{blaizot1986} . This self-consistent mean-field method was used previously for two-dimensional dipolar fermions~\cite{bhongale2012,bhongale2013}. 
Although anomalous mean-field terms that allow for a description of pairing with arbitrary spatial symmetry can in principle be included in the self-consistent method, we set them to zero in this work.
As can be seen in Fig.~\ref{fig:occ_eq}, both L-DMFT and DMFT give qualitatively similar results to the self-consistent mean-field method, which shows that terms beyond Hartree approximation in the nearest-neighbor interaction do not modify qualitatively the density-wave ordering in this system. The discrepancies at larger values of $U$ are expected to be an effect of mean-field treatment of local interactions rather than neglecting exchange terms in DMFT or L-DMFT.

\subsection{Non-equilibrium results}
\label{sec:noneq}

\begin{figure}[pt!]
\begin{center}
\resizebox{1.\columnwidth}{!}{
\includegraphics{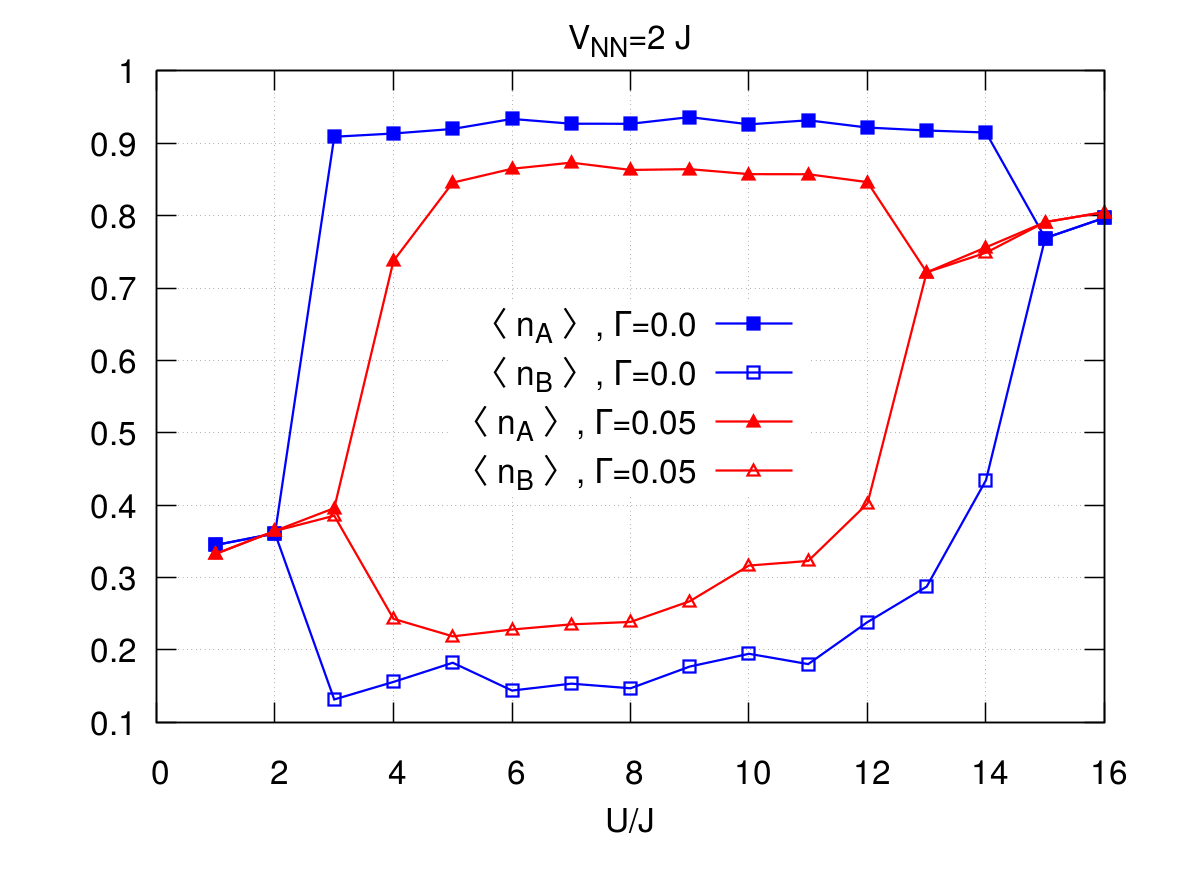}}
\caption{\label{fig:gm0_vs_gm05} Comparison of the occupations of sublattice $A$ (filled symbols) and $B$ (empty symbols) for no dephasing $\Gamma_\mathrm{dp}=0.0$ (blue squares) and intermediate dephasing strength $\Gamma_\mathrm{dp}=0.05$ (red triangles).}
\end{center}
\end{figure}%

\begin{figure}[pt!]
\begin{center}
\resizebox{1.\columnwidth}{!}{
\includegraphics{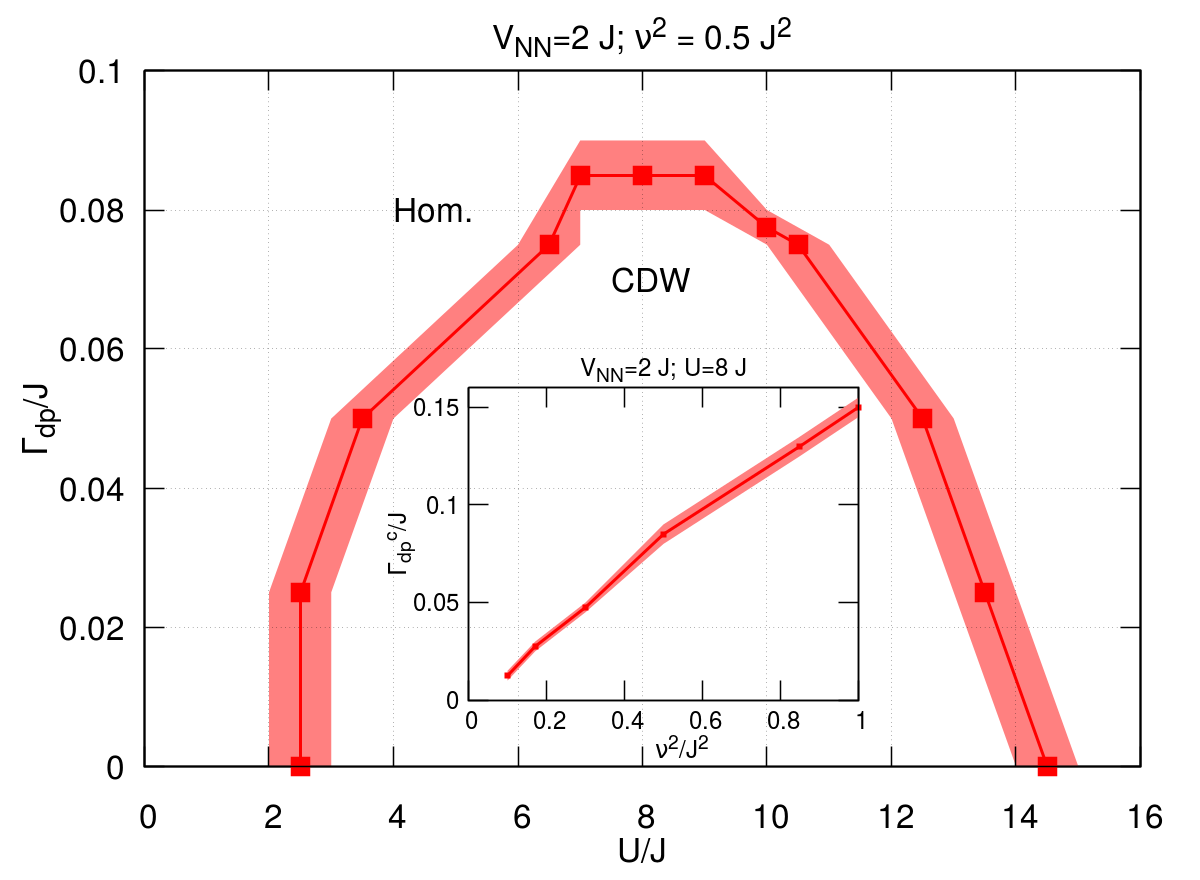}}
\caption{\label{fig:phase_diag} Estimated phase diagram as a function of on-site interaction $U$ and dephasing $\Gamma_\mathrm{dp}$. Other parameters are set to $\nu^2=0.5$, $V=2$, $\epsilon=-U/2$. Shaded region represents accuracy with which we determine the phase transition line between CDW and homogeneous phases. Inset: Critical value of $\Gamma_\mathrm{dp}$ at constant $U=8J$ as a function of $\nu^2$.}
\end{center}
\end{figure}%

We now turn to the non-equilibrium case. In Fig. \ref{fig:gm0_vs_gm05} we compare the occupations of sublattices $A$ and $B$ for different interaction strengths with and without the dephasing $\Gamma_\mathrm{dp}=0.05$. We observe that the dephasing has the effect of reducing the differences between the occupation of sublattice $A$ ($\langle\hat{n}_A\rangle$) and sublattice $B$ ($\langle\hat{n}_B\rangle$). We also observe a change of the critical value of the local interaction, at which the system undergoes a phase transition between the homogeneous and CDW phases. The range of values of $U$ for which the CDW phase is present thus becomes smaller due to dephasing.


In order to further estimate the destructive effect of dephasing we investigate the $U-\Gamma_\mathrm{dp}$ phase diagram of our system, Fig. \ref{fig:phase_diag}. It is evident that with increasing strength of dephasing the range in $U$ for which one obtains a CDW phase shrinks, until it vanishes completely at around $\Gamma_\mathrm{dp}\approx 0.085\pm0.005$. Note that this value is  at least one order of magnitude smaller than the other energy scales of the system, $\epsilon$, $U$, $V$ and $J$. 
We also note that the $U$-dependence of the critical value of $\Gamma_\mathrm{dp}$ is approximately symmetric with respect to the maximum at around $U=8$.

Next, we investigate how the coupling strength to the local thermal baths $\nu^2$ affects the results. In the inset of Fig.~\ref{fig:phase_diag} we present the dependence of the critical dephasing strength $\Gamma^c_\mathrm{dp}$ on this quantity for $U=8$. 
We observe that for $\nu^2\lessapprox0.5$ the two quantities are proportional to each other $\Gamma^c_\mathrm{dp}\sim\nu^2$, whereas at large $\nu^2$ the critical value of $\Gamma^c_\mathrm{dp}$ seems to be shifted away from proportionality to lower values.

One possible mechanism to explain the proportionality for small $\nu^2$ is the presence of heat exchange between the lattice and the baths, as we expect the heat current in lattice systems to be proportional to $\nu^2$~\cite{titvinidze2017}. In this regime the critical value of the dephasing would then be determined by the rate at which the heat generated by the dephasing is taken out of the system. 

At high $\nu^2$ we expect this behavior to change because of the increasing hybridization between the system and the local thermal baths. In this regime, the baths have a stronger effect on the system. Although the rate of cooling is faster, at the same time they do not favor CDW ordering since they are identical for both sublattices. However, whether this is the only mechanism affecting the behavior at high $\nu^2$ cannot be concluded.

In the cold-atom experiments the cooling rate is not easy to control (though not impossible~\cite{chong2018}). In the absence of controllable coupling to a thermal bath, however, one could still observe experimentally the CDW phase if the time scale at which the system is heated by dissipation is much longer than the time scale at which the CDW ordering emerges.

To check whether we work with experimentally realistic physical parameters, we consider now a possible experimental realization with fermionic $^{40}$K atoms loaded into an optical lattice~\cite{joerdens2008,schneider2008}. As discussed in Sec. \ref{sec:ham} such a system can be well described by the Hubbard model and in order to introduce non-local interactions one can couple the two spin states to highly excited Rydberg states in the weak dressing regime, see also App. \ref{appsec:dress}. For sufficiently large detuning compared to the Rabi frequency $\delta\gg\Omega$ one obtains effectively dressed ground states with Eq.~\eqref{eq:VvdWef} describing effective interaction potential of two atoms~\cite{henkel2010}. Here the effective coefficient is given by $\tilde{C}_6=(\Omega/2\delta)^4 C_6$, with $C_6$ determining the strength of the van der Waals interactions between two Rydberg states. The soft-core radius is given by $R_c=(C_6/2|\delta|)^{1/6}$. Finally, the dephasing strength is determined via the spontaneous emission rate $\Gamma_\mathrm{se}$ of the excited state via $\Gamma_\mathrm{dp} = (\Omega/2\delta)^2\Gamma_\mathrm{se}$, see App. \ref{appsec:dress}.

Let us now consider a particular choice of the Rydberg state for the $^{40}$K atoms, namely the $|26S\rangle$ state. For this choice we obtain $C_6\approx 27\ \hbar\ \mathrm{MHz}\ \mu\mathrm{m}^6$~\cite{singer2005}, and $\Gamma_\mathrm{se}\approx 60\ \hbar\ $kHz~\cite{beterov2009}. A typical value of the hopping amplitude in optical lattice is on the order of $J\approx 0.5\ \hbar\ $kHz and the lattice spacing is on the order of $a\approx 0.5\mu$m.

We aim at realizing a model in which only nearest-neighbor interaction is relevant. Therefore, we set the parameters to $R_c=0.5\mu$m and $\tilde{C}_6= 31\ \hbar\ \mathrm{Hz}\ \mu\mathrm{m}^6$. In this case we would obtain a nearest-neighbor interaction strength $V\approx 2J$, the same as in our calculations, and a next-nearest-neighbor interaction which is at least one order of magnitude weaker. To obtain the required value of $R_c$ one needs to set the detuning on the order of $\delta\approx860\ \hbar\ $MHz. With this value of the detuning and in order to get the appropriate value of $\tilde{C}_6$, we need to set the Rabi frequency to $\Omega=56\ \hbar\ $MHz (suggesting the need for further development of current experimental capabilities). Finally, we use the values of $\Omega$ and $\delta$ to estimate the effective dephasing strength, which is approximately given by $\Gamma_\mathrm{dp}\approx 64 \ \hbar\ $Hz $=0.128J$ -- on the order of magnitude of the maximal $\Gamma_\mathrm{dp}$ considered here. The time scale at which dephasing heats the system is approximately given by $\hbar/\Gamma_\mathrm{dp}\approx 16 ms$.

We note that there are several ways in which one can decrease the dephasing strength in experiment, e.g., (i) with higher values of the Rabi frequency one can target higher excited states, which have a longer lifetime, (ii) one can use a lattice with a larger lattice constant, which also allows to use higher excited states without increasing the Rabi frequency.
Note, however, that in our analysis we have neglected effects of black body radiation, which are present in a typical experiment and can lead to both stronger dephasing and avalanche loss of particles from the system~\cite{goldschmidt2016,zeiher2016}. 

\section{Conclusions}
\label{sec:end}
In this work we have studied the effect of dissipation on charge ordered density-wave phases in a strongly-correlated many-body quantum system with local and non-local interaction, as encompassed by the fermionic extended Hubbard model, with dissipation effects treated at the level of the quantum master equation. This model was solved using a recent variant of non-equilibrium dynamical mean-field theory, the Lindblad-DMFT, that allows to include local dissipation effects non-perturbatively. 

By studying the behavior of the checkerboard CDW order parameter, we have demonstrated that a CDW phase, similar to the one present in the zero-temperature equilibrium model, survives the introduction of a dephasing process up to a critical strength, where the density ordering is destroyed and the system becomes homogeneous. We studied the steady-state phase diagram of the model as a function of the local interaction $U$ and dissipation strength $\Gamma_\mathrm{dp}$ and found that a broad region of density-ordered steady-states exists at relatively weak and moderate dephasing strengths.
We observed that the critical value of local interaction $U$, where the phase transition between the homogeneous and CDW phases occurs, depends on the dephasing strength, with the CDW phase shrinking as the dephasing strength is increased.
Importantly, we observed that to a certain extent the effect of dephasing on the CDW order seems to be due dominantly to heating, as we have observed that the critical value of $\Gamma^c_\mathrm{dp}$ is proportional to the coupling strength $\nu^2$ to the bath.

We expect that using cold atomic fermionic gases dressed with a Rydberg state -- thus acquiring long-range interactions -- and loaded into optical lattices could present an experimental realization of the extended Hubbard model. We showed that the parameters considered in our work are experimentally realistic. The remaining issue is to estimate the effect of other types of dissipation, and estimate the time scale at which CDW order emerges in such a system and make sure that it is much shorter then the time scale at which the system is heated by dephasing.

\begin{acknowledgments}
Support by the Deutsche Forschungsgemeinschaft via DFG SPP 1929 GiRyd, SFB/TR 49 and the high-performance computing center LOEWE-CSC is gratefully acknowledged. I.T. and M. S. acknowledge funding from the Austrian Science Fund (FWF) within Projects P26508 and F41 (SFB ViCoM). The authors also acknowledge useful discussions with K. Byczuk, C. Gro\ss{}, H. Weimer, S. Whitlock and J. Zeiher.
\end{acknowledgments}

\appendix

\section{Dressed regime}
\label{appsec:dress}
In order to simulate a spin-$\frac{1}{2}$ fermionic Hubbard model one can use two hyperfine states of $^{40}$K potassium atoms~\cite{joerdens2008,schneider2008}. To introduce the long range interaction we need to couple these states to high lying Rydberg excited states. Because we want the non-local interaction to be isotropic we either need to use $|nS\rangle$ Rydberg states with a three level excitation scheme~\cite{goldschmidt2016,aman2016} or $|nP\rangle$ Rydberg states with a two level excitation scheme~\cite{zeiher2016}, but then one needs to appropriately arrange the orientation of Rydberg states with respect to the 2D lattice. In both cases it is enough to work with a single effective Rabi frequency $\Omega_{eff}$ and detuning $\delta_{eff}$.

Because we need to couple two hyperfine states to excited Rydberg states one can choose either to couple both hyperfine states to the same Rydberg state, or to couple each hyperfine state to a different Rydberg state, as depicted in Fig. \ref{fig:model}. The first approach gives the same inter- and intra-species non-local interaction, but introduces small coherent and incoherent spin flip processes. The second approach does not introduce spin-flip terms, but results in small differences in the non-local interaction strength of different species. In the following we assume that these differences are negligible.

The full model for the corresponding experimental set-up, in the rotating wave approximation, has the following Hamiltonian~\cite{saha2014}
\begin{equation}
\label{apeq:mod}
\begin{split}
\hat{H}_1=&\sum_{\sigma\in\{\uparrow,\downarrow\}} \sum_{\langle i, j \rangle} \left(-J\hat{f}^\dagger_{i,\sigma}\hat{f}^{\phantom\dag}_{j,\sigma} -\tilde{J}\hat{f}^\dagger_{i,R_\sigma}\hat{f}^{\phantom\dag}_{j,R_\sigma}\right) \\
&  +\frac{\Omega_{eff}}{2} \sum_{i,\sigma\in\{\uparrow,\downarrow\}} \left(\hat{f}^\dagger_{i,\sigma}\hat{f}^{\phantom\dag}_{i,R_\sigma}+\mathrm{h.c.}\right) \\ 
&-\delta_{eff} \sum_{i,\sigma} \hat{n}_{i,R_\sigma} + U \sum_i \hat{n}_{i, \uparrow}\hat{n}_{i, \downarrow} \\
& + \sum_{i, j, \sigma, \sigma'}\frac{V_{RR}(\textbf{r}_i,\textbf{r}_j)}{2} \hat{n}_{i,R_\sigma}  \hat{n}_{j,R_{\sigma'}} \\
& + \sum_{i, j,\sigma,\sigma'}V_{gR}(\textbf{r}_i,\textbf{r}_j) \hat{n}_{i,\sigma}  \hat{n}_{j,R_{\sigma'}}.
\end{split}
\end{equation}
Here, apart from terms appearing in the Eq.~\eqref{eq:ehm}, we have the hopping amplitude $\tilde{J}$ of the excited states, the effective Rabi frequency $\Omega_{eff}$ and detuning $\delta_{eff}$, and non-local interaction strengths $V_{RR}(\textbf{r}_i,\textbf{r}_j)$, $V_{gR}(\textbf{r}_i,\textbf{r}_j)$. $\hat{f}_{i,\sigma}$ annihilates a ground state atom in a hyperfine state $\sigma$ on site $i$. We use the notation $\sigma\in\{\uparrow,\downarrow\}$ for the two hyperfine states as they are later interpreted as two spin states of the Hubbard model. $\hat{f}_{i,R_\sigma}$ annihilates on site $i$ an atom in an excited Rydberg state $R_\sigma$, to which the hyperfine state $\sigma$ is coupled, see Fig.~\ref{fig:model}.

As the system is subject to dissipative processes, it is not enough to determine the Hamiltonian, but we also need to determine the Lindblad operators. Here we will consider only spontaneous emission. The Lindblad operator for the spontaneous emission is
\begin{equation}
\label{apeq:spem}
\hat{L}^{se}_{i,\sigma} = \hat{f}^{\dag}_{i,\sigma}\hat{f}^{\phantom\dag}_{i,R_\sigma}
\end{equation}
with the strength of dissipation given by the constant $\Gamma_\mathrm{se}$ independent of spin and position. When coupling to the $|nS\rangle$ state via a 3-level scheme~\cite{goldschmidt2016,aman2016}, this form of the Lindblad operator is approximate, assuming that the decay of the atom to any intermediate state is immediately followed by decay to the ground state.

For the moment we consider a single atom without dephasing. Due to the Rabi driving its ground and excited states are no longer eigenstates of the full Hamiltonian. E.g., the lowest energy eigenstate has the form $|\tilde{\sigma}\rangle = \alpha|\sigma\rangle + \beta |R_{\sigma}\rangle$. Assuming that we are in the regime where $\delta_{eff}\gg\Omega_{eff}$, we have that $\alpha\approx1$ and $\beta\approx \Omega_{eff}/(2\delta_{eff})\ll 1$~\cite{zeiher2016}. The admixture of the excited state to the new eigenstate is small. The state $|\tilde{\sigma}\rangle$ is called the dressed ground state. Similarly, the dressed high energy eigenstate $|\tilde{R}_\sigma\rangle$ will be predominantly a Rydberg state, with a small admixture of the ground state. It is safe to assume that due to its high energy, the $|\tilde{R}_\sigma\rangle$ is empty, and only the dressed ground state is occupied.

Next we consider two atoms without dephasing. Due to the small admixture of Rydberg excitation to the dressed ground state, and due to the very strong non-local interaction between two Rydberg states, the dressed ground state will effectively be subject to non-local interaction. The strength of this interaction will be approximately proportional to $V_{eff}\approx\beta^4 V_{RR}$~\cite{zeiher2016}. This interaction shifts the atom even further from resonance and as a results the value of $\beta$ will depend on the distance separating two atoms, therefore the effective interaction will also have a renormalized shape, with a soft core cut-off for small atom-atom separations. In this way we get to the Eq.~\eqref{eq:VvdWef} for the effective potential~\cite{henkel2010}.

Note that through appropriate choice of $\Omega_{eff}$ and $\delta_{eff}$ we can control the strength and shape of the interaction potential. Thanks to this flexibility we can set the parameters such that only nearest-neighbor interaction is relevant in our model. Using the notation in which $\hat{c}_{i,\sigma}$ and $\hat{c}_{i,R_\sigma}$ annihilate the dressed ground and excited states, respectively, corresponding to spin $\sigma$ on site $i$, we obtain the Hamiltonian \eqref{eq:ehm}.

Finally, we consider the effect of dissipation in the dressed regime. The operators corresponding to the dressed states can be written as $\hat{f}_{i,\sigma} = \alpha^\ast \hat{c}_{i,\sigma} - \beta \hat{c}_{i,R_\sigma}$ and $\hat{f}_{i,R_\sigma} = \beta^{\ast} \hat{c}_{i,\sigma} + \alpha \hat{c}_{i,R_\sigma}$ with $|\alpha|^2+|\beta|^2=1$. In this representation the Lindblad operator \eqref{apeq:spem} becomes 
\begin{equation}
\hat{L}^{se}_{i,\sigma} = \left(\alpha^\ast \hat{c}^\dag_{i,\sigma} - \beta \hat{c}^\dag_{i,R_\sigma}\right) \left(\beta^\ast \hat{c}^{\phantom\dag}_{i,\sigma} + \alpha \hat{c}^{\phantom\dag}_{i,R_\sigma}\right).
\end{equation}
Under conditions in which $\alpha\approx 1$ and $\beta\ll1$ the term with pre-factor $\beta^2$ will vanish. As we also mentioned the dressed excited state is empty, as it is a high-energy eigenstate. Therefore terms with $\hat{c}^{\phantom\dag}_{i,R_\sigma}$ can also be neglected. One finally obtains 
\begin{equation}
\hat{L}^{se}_{i,\sigma} \approx \beta^\ast \hat{c}^\dag_{i,\sigma}  \hat{c}^{\phantom\dag}_{i,\sigma}.
\end{equation}
We obtain effectively a new type of dissipation, namely dephasing, with Lindblad operator given by \eqref{eq:Ldp} and dissipation strength given by $\Gamma_\mathrm{dp}=|\beta|^2\Gamma_\mathrm{se}$. In this picture the dressed excited states corresponding to $\hat{c}_{i,R_\sigma}$ drop out completely.

\section{Dissipative heating}
\label{appsec:disheat}

To show that the dephasing considered in section \ref{sec:ham} indeed heats the system, we consider a case in which our model \eqref{eq:ehm} is decoupled from thermal baths and the dissipation strength is finite. We notice that the Liouvillian \eqref{eq:liouvillian2} with Lindblad operator \eqref{eq:Ldp} conserves the number of particles in the system for both spin species. Therefore, we will consider states with fixed number of particles at half-filling. In other cases the Fock space of the system can be split into subspaces with different occupations and analyzed separately. 

The maximally mixed state of a system corresponds to an infinite temperature. We will show that such a state is indeed a steady-state of the system. The uniqueness of the steady-state~\cite{kraus2008,spohn1976,spohn1977} has been studied for certain specific classes of Liouvillians, but we have not found a proof applicable in our case. Nevertheless, we will assume that within the subspace of fixed particle number the steady-state is unique. With this one can conclude that dephasing heats up the system.

The maximally mixed state within the Fock space of dimension $M$ is proportional to the identity operator
\begin{equation}
\hat{\rho}_{MM}=\frac{1}{M} \hat{\mathbb{1}}.
\end{equation}
The condition for it to be a steady-state reads
\begin{equation}
\frac{d \hat{\rho}_{MM}}{d t} = -i\left[\hat{H}, \hat{\rho}_{MM} \right] + \doublehat{\mathcal{L}}[\hat{\rho}_{MM}] = 0
\end{equation}
As any operator commutes with the identity operator, the first term on the right-hand side vanishes. Now we consider the dissipative part given by \eqref{eq:liouvillian2} with the Lindblad jump operator \eqref{eq:Ldp} and $\Gamma_{\mu,\nu}=\delta_{\mu,\nu} \Gamma_\mathrm{dp}$, where $\mu$ iterates over lattice $i$ and spin $\sigma$ indices. Notice that as the jump operator is hermitian we get
\begin{equation}
2 \hat{L}^{\phantom\dag}_{dp,i,\sigma} \hat{\mathbb{1}} \hat{L}^\dag_{dp,i,\sigma} - \left\{\hat{L}^\dag_{dp,i,\sigma}\hat{L}^{\phantom\dag}_{dp,i,\sigma},\hat{\mathbb{1}} \right\} =0.
\end{equation}
As a result, the time derivative of $\hat{\rho}_{MM}$ vanishes, which means that it is indeed a steady-state.

\bibliography{main_text_file}

\begin{thebibliography}{92}%
\makeatletter
\providecommand \@ifxundefined [1]{%
 \@ifx{#1\undefined}
}%
\providecommand \@ifnum [1]{%
 \ifnum #1\expandafter \@firstoftwo
 \else \expandafter \@secondoftwo
 \fi
}%
\providecommand \@ifx [1]{%
 \ifx #1\expandafter \@firstoftwo
 \else \expandafter \@secondoftwo
 \fi
}%
\providecommand \natexlab [1]{#1}%
\providecommand \enquote  [1]{``#1''}%
\providecommand \bibnamefont  [1]{#1}%
\providecommand \bibfnamefont [1]{#1}%
\providecommand \citenamefont [1]{#1}%
\providecommand \href@noop [0]{\@secondoftwo}%
\providecommand \href [0]{\begingroup \@sanitize@url \@href}%
\providecommand \@href[1]{\@@startlink{#1}\@@href}%
\providecommand \@@href[1]{\endgroup#1\@@endlink}%
\providecommand \@sanitize@url [0]{\catcode `\\12\catcode `\$12\catcode
  `\&12\catcode `\#12\catcode `\^12\catcode `\_12\catcode `\%12\relax}%
\providecommand \@@startlink[1]{}%
\providecommand \@@endlink[0]{}%
\providecommand \url  [0]{\begingroup\@sanitize@url \@url }%
\providecommand \@url [1]{\endgroup\@href {#1}{\urlprefix }}%
\providecommand \urlprefix  [0]{URL }%
\providecommand \Eprint [0]{\href }%
\providecommand \doibase [0]{http://dx.doi.org/}%
\providecommand \selectlanguage [0]{\@gobble}%
\providecommand \bibinfo  [0]{\@secondoftwo}%
\providecommand \bibfield  [0]{\@secondoftwo}%
\providecommand \translation [1]{[#1]}%
\providecommand \BibitemOpen [0]{}%
\providecommand \bibitemStop [0]{}%
\providecommand \bibitemNoStop [0]{.\EOS\space}%
\providecommand \EOS [0]{\spacefactor3000\relax}%
\providecommand \BibitemShut  [1]{\csname bibitem#1\endcsname}%
\let\auto@bib@innerbib\@empty
\bibitem [{\citenamefont {Goldschmidt}\ \emph {et~al.}(2016)\citenamefont
  {Goldschmidt}, \citenamefont {Boulier}, \citenamefont {Brown}, \citenamefont
  {Koller}, \citenamefont {Young}, \citenamefont {Gorshkov}, \citenamefont
  {Rolston},\ and\ \citenamefont {Porto}}]{goldschmidt2016}%
  \BibitemOpen
  \bibfield  {author} {\bibinfo {author} {\bibfnamefont {E.~A.}\ \bibnamefont
  {Goldschmidt}}, \bibinfo {author} {\bibfnamefont {T.}~\bibnamefont
  {Boulier}}, \bibinfo {author} {\bibfnamefont {R.~C.}\ \bibnamefont {Brown}},
  \bibinfo {author} {\bibfnamefont {S.~B.}\ \bibnamefont {Koller}}, \bibinfo
  {author} {\bibfnamefont {J.~T.}\ \bibnamefont {Young}}, \bibinfo {author}
  {\bibfnamefont {A.~V.}\ \bibnamefont {Gorshkov}}, \bibinfo {author}
  {\bibfnamefont {S.~L.}\ \bibnamefont {Rolston}}, \ and\ \bibinfo {author}
  {\bibfnamefont {J.~V.}\ \bibnamefont {Porto}},\ }\bibfield  {title} {\enquote
  {\bibinfo {title} {Anomalous broadening in driven dissipative rydberg
  systems},}\ }\href {\doibase 10.1103/PhysRevLett.116.113001} {\bibfield
  {journal} {\bibinfo  {journal} {Phys. Rev. Lett.}\ }\textbf {\bibinfo
  {volume} {116}},\ \bibinfo {pages} {113001} (\bibinfo {year}
  {2016})}\BibitemShut {NoStop}%
\bibitem [{\citenamefont {Aman}\ \emph {et~al.}(2016)\citenamefont {Aman},
  \citenamefont {DeSalvo}, \citenamefont {Dunning}, \citenamefont {Killian},
  \citenamefont {Yoshida},\ and\ \citenamefont {Burgd\"orfer}}]{aman2016}%
  \BibitemOpen
  \bibfield  {author} {\bibinfo {author} {\bibfnamefont {J.~A.}\ \bibnamefont
  {Aman}}, \bibinfo {author} {\bibfnamefont {B.~J.}\ \bibnamefont {DeSalvo}},
  \bibinfo {author} {\bibfnamefont {F.~B.}\ \bibnamefont {Dunning}}, \bibinfo
  {author} {\bibfnamefont {T.~C.}\ \bibnamefont {Killian}}, \bibinfo {author}
  {\bibfnamefont {S.}~\bibnamefont {Yoshida}}, \ and\ \bibinfo {author}
  {\bibfnamefont {J.}~\bibnamefont {Burgd\"orfer}},\ }\bibfield  {title}
  {\enquote {\bibinfo {title} {Trap losses induced by near-resonant rydberg
  dressing of cold atomic gases},}\ }\href {\doibase
  10.1103/PhysRevA.93.043425} {\bibfield  {journal} {\bibinfo  {journal} {Phys.
  Rev. A}\ }\textbf {\bibinfo {volume} {93}},\ \bibinfo {pages} {043425}
  (\bibinfo {year} {2016})}\BibitemShut {NoStop}%
\bibitem [{\citenamefont {Zeiher}\ \emph {et~al.}(2015)\citenamefont {Zeiher},
  \citenamefont {Schau\ss{}}, \citenamefont {Hild}, \citenamefont {Macr\`{\i}},
  \citenamefont {Bloch},\ and\ \citenamefont {Gross}}]{zeiher2015}%
  \BibitemOpen
  \bibfield  {author} {\bibinfo {author} {\bibfnamefont {J.}~\bibnamefont
  {Zeiher}}, \bibinfo {author} {\bibfnamefont {P.}~\bibnamefont {Schau\ss{}}},
  \bibinfo {author} {\bibfnamefont {S.}~\bibnamefont {Hild}}, \bibinfo {author}
  {\bibfnamefont {T.}~\bibnamefont {Macr\`{\i}}}, \bibinfo {author}
  {\bibfnamefont {I.}~\bibnamefont {Bloch}}, \ and\ \bibinfo {author}
  {\bibfnamefont {C.}~\bibnamefont {Gross}},\ }\bibfield  {title} {\enquote
  {\bibinfo {title} {Microscopic characterization of scalable coherent rydberg
  superatoms},}\ }\href {\doibase 10.1103/PhysRevX.5.031015} {\bibfield
  {journal} {\bibinfo  {journal} {Phys. Rev. X}\ }\textbf {\bibinfo {volume}
  {5}},\ \bibinfo {pages} {031015} (\bibinfo {year} {2015})}\BibitemShut
  {NoStop}%
\bibitem [{\citenamefont {Zeiher}\ \emph {et~al.}(2016)\citenamefont {Zeiher},
  \citenamefont {Van~Bijnen}, \citenamefont {Schau{\ss}}, \citenamefont {Hild},
  \citenamefont {Choi}, \citenamefont {Pohl}, \citenamefont {Bloch},\ and\
  \citenamefont {Gross}}]{zeiher2016}%
  \BibitemOpen
  \bibfield  {author} {\bibinfo {author} {\bibfnamefont {J.}~\bibnamefont
  {Zeiher}}, \bibinfo {author} {\bibfnamefont {R.}~\bibnamefont {Van~Bijnen}},
  \bibinfo {author} {\bibfnamefont {P.}~\bibnamefont {Schau{\ss}}}, \bibinfo
  {author} {\bibfnamefont {S.}~\bibnamefont {Hild}}, \bibinfo {author}
  {\bibfnamefont {J.}~\bibnamefont {Choi}}, \bibinfo {author} {\bibfnamefont
  {T.}~\bibnamefont {Pohl}}, \bibinfo {author} {\bibfnamefont {I.}~\bibnamefont
  {Bloch}}, \ and\ \bibinfo {author} {\bibfnamefont {C.}~\bibnamefont
  {Gross}},\ }\bibfield  {title} {\enquote {\bibinfo {title} {Many-body
  interferometry of a rydberg-dressed spin lattice},}\ }\href
  {http://dx.doi.org/10.1038/nphys3835} {\bibfield  {journal} {\bibinfo
  {journal} {Nat. Phys.}\ }\textbf {\bibinfo {volume} {12}},\ \bibinfo {pages}
  {1095} (\bibinfo {year} {2016})}\BibitemShut {NoStop}%
\bibitem [{\citenamefont {Vidanovi\ifmmode~\acute{c}\else \'{c}\fi{}}\ \emph
  {et~al.}(2014)\citenamefont {Vidanovi\ifmmode~\acute{c}\else \'{c}\fi{}},
  \citenamefont {Cocks},\ and\ \citenamefont {Hofstetter}}]{vidanovic2014}%
  \BibitemOpen
  \bibfield  {author} {\bibinfo {author} {\bibfnamefont {I.}~\bibnamefont
  {Vidanovi\ifmmode~\acute{c}\else \'{c}\fi{}}}, \bibinfo {author}
  {\bibfnamefont {D.}~\bibnamefont {Cocks}}, \ and\ \bibinfo {author}
  {\bibfnamefont {W.}~\bibnamefont {Hofstetter}},\ }\bibfield  {title}
  {\enquote {\bibinfo {title} {Dissipation through localized loss in bosonic
  systems with long-range interactions},}\ }\href {\doibase
  10.1103/PhysRevA.89.053614} {\bibfield  {journal} {\bibinfo  {journal} {Phys.
  Rev. A}\ }\textbf {\bibinfo {volume} {89}},\ \bibinfo {pages} {053614}
  (\bibinfo {year} {2014})}\BibitemShut {NoStop}%
\bibitem [{\citenamefont {Bernier}\ \emph {et~al.}(2014)\citenamefont
  {Bernier}, \citenamefont {Poletti},\ and\ \citenamefont
  {Kollath}}]{bernier2014}%
  \BibitemOpen
  \bibfield  {author} {\bibinfo {author} {\bibfnamefont {J.-S.}\ \bibnamefont
  {Bernier}}, \bibinfo {author} {\bibfnamefont {D.}~\bibnamefont {Poletti}}, \
  and\ \bibinfo {author} {\bibfnamefont {C.}~\bibnamefont {Kollath}},\
  }\bibfield  {title} {\enquote {\bibinfo {title} {Dissipative quantum dynamics
  of fermions in optical lattices: A slave-spin approach},}\ }\href {\doibase
  10.1103/PhysRevB.90.205125} {\bibfield  {journal} {\bibinfo  {journal} {Phys.
  Rev. B}\ }\textbf {\bibinfo {volume} {90}},\ \bibinfo {pages} {205125}
  (\bibinfo {year} {2014})}\BibitemShut {NoStop}%
\bibitem [{\citenamefont {Sarkar}\ \emph {et~al.}(2014)\citenamefont {Sarkar},
  \citenamefont {Langer}, \citenamefont {Schachenmayer},\ and\ \citenamefont
  {Daley}}]{sarkar2014}%
  \BibitemOpen
  \bibfield  {author} {\bibinfo {author} {\bibfnamefont {S.}~\bibnamefont
  {Sarkar}}, \bibinfo {author} {\bibfnamefont {S.}~\bibnamefont {Langer}},
  \bibinfo {author} {\bibfnamefont {J.}~\bibnamefont {Schachenmayer}}, \ and\
  \bibinfo {author} {\bibfnamefont {A.~J.}\ \bibnamefont {Daley}},\ }\bibfield
  {title} {\enquote {\bibinfo {title} {Light scattering and dissipative
  dynamics of many fermionic atoms in an optical lattice},}\ }\href {\doibase
  10.1103/PhysRevA.90.023618} {\bibfield  {journal} {\bibinfo  {journal} {Phys.
  Rev. A}\ }\textbf {\bibinfo {volume} {90}},\ \bibinfo {pages} {023618}
  (\bibinfo {year} {2014})}\BibitemShut {NoStop}%
\bibitem [{\citenamefont {Kraus}\ \emph {et~al.}(2008)\citenamefont {Kraus},
  \citenamefont {B\"uchler}, \citenamefont {Diehl}, \citenamefont {Kantian},
  \citenamefont {Micheli},\ and\ \citenamefont {Zoller}}]{kraus2008}%
  \BibitemOpen
  \bibfield  {author} {\bibinfo {author} {\bibfnamefont {B.}~\bibnamefont
  {Kraus}}, \bibinfo {author} {\bibfnamefont {H.~P.}\ \bibnamefont
  {B\"uchler}}, \bibinfo {author} {\bibfnamefont {S.}~\bibnamefont {Diehl}},
  \bibinfo {author} {\bibfnamefont {A.}~\bibnamefont {Kantian}}, \bibinfo
  {author} {\bibfnamefont {A.}~\bibnamefont {Micheli}}, \ and\ \bibinfo
  {author} {\bibfnamefont {P.}~\bibnamefont {Zoller}},\ }\bibfield  {title}
  {\enquote {\bibinfo {title} {Preparation of entangled states by quantum
  markov processes},}\ }\href {\doibase 10.1103/PhysRevA.78.042307} {\bibfield
  {journal} {\bibinfo  {journal} {Phys. Rev. A}\ }\textbf {\bibinfo {volume}
  {78}},\ \bibinfo {pages} {042307} (\bibinfo {year} {2008})}\BibitemShut
  {NoStop}%
\bibitem [{\citenamefont {Diehl}\ \emph {et~al.}(2008)\citenamefont {Diehl},
  \citenamefont {Micheli}, \citenamefont {Kantian}, \citenamefont {Kraus},
  \citenamefont {B\"uchler},\ and\ \citenamefont {Zoller}}]{diehl2008}%
  \BibitemOpen
  \bibfield  {author} {\bibinfo {author} {\bibfnamefont {S.}~\bibnamefont
  {Diehl}}, \bibinfo {author} {\bibfnamefont {A.}~\bibnamefont {Micheli}},
  \bibinfo {author} {\bibfnamefont {A.}~\bibnamefont {Kantian}}, \bibinfo
  {author} {\bibfnamefont {B.}~\bibnamefont {Kraus}}, \bibinfo {author}
  {\bibfnamefont {H.~P.}\ \bibnamefont {B\"uchler}}, \ and\ \bibinfo {author}
  {\bibfnamefont {P.}~\bibnamefont {Zoller}},\ }\bibfield  {title} {\enquote
  {\bibinfo {title} {Quantum states and phases in driven open quantum systems
  with cold atoms},}\ }\href {http://dx.doi.org/10.1038/nphys1073} {\bibfield
  {journal} {\bibinfo  {journal} {Nat. Phys.}\ }\textbf {\bibinfo {volume}
  {4}},\ \bibinfo {pages} {878} (\bibinfo {year} {2008})}\BibitemShut {NoStop}%
\bibitem [{\citenamefont {Dzhioev}\ and\ \citenamefont
  {Kosov}(2011{\natexlab{a}})}]{dzhioev2011}%
  \BibitemOpen
  \bibfield  {author} {\bibinfo {author} {\bibfnamefont {A.~A.}\ \bibnamefont
  {Dzhioev}}\ and\ \bibinfo {author} {\bibfnamefont {D.~S.}\ \bibnamefont
  {Kosov}},\ }\bibfield  {title} {\enquote {\bibinfo {title} {Stability
  analysis of multiple nonequilibrium fixed points in self-consistent electron
  transport calculations},}\ }\href {\doibase 10.1063/1.3658736} {\bibfield
  {journal} {\bibinfo  {journal} {The Journal of Chemical Physics}\ }\textbf
  {\bibinfo {volume} {135}},\ \bibinfo {pages} {174111} (\bibinfo {year}
  {2011}{\natexlab{a}})},\ \Eprint
  {http://arxiv.org/abs/https://doi.org/10.1063/1.3658736}
  {https://doi.org/10.1063/1.3658736} \BibitemShut {NoStop}%
\bibitem [{\citenamefont {Dorda}\ \emph {et~al.}(2014)\citenamefont {Dorda},
  \citenamefont {Nuss}, \citenamefont {von~der Linden},\ and\ \citenamefont
  {Arrigoni}}]{dorda2014}%
  \BibitemOpen
  \bibfield  {author} {\bibinfo {author} {\bibfnamefont {A.}~\bibnamefont
  {Dorda}}, \bibinfo {author} {\bibfnamefont {M.}~\bibnamefont {Nuss}},
  \bibinfo {author} {\bibfnamefont {W.}~\bibnamefont {von~der Linden}}, \ and\
  \bibinfo {author} {\bibfnamefont {E.}~\bibnamefont {Arrigoni}},\ }\bibfield
  {title} {\enquote {\bibinfo {title} {Auxiliary master equation approach to
  nonequilibrium correlated impurities},}\ }\href {\doibase
  10.1103/PhysRevB.89.165105} {\bibfield  {journal} {\bibinfo  {journal} {Phys.
  Rev. B}\ }\textbf {\bibinfo {volume} {89}},\ \bibinfo {pages} {165105}
  (\bibinfo {year} {2014})}\BibitemShut {NoStop}%
\bibitem [{\citenamefont {Schwarz}\ \emph {et~al.}(2016)\citenamefont
  {Schwarz}, \citenamefont {Goldstein}, \citenamefont {Dorda}, \citenamefont
  {Arrigoni}, \citenamefont {Weichselbaum},\ and\ \citenamefont {von
  Delft}}]{schwarz2016}%
  \BibitemOpen
  \bibfield  {author} {\bibinfo {author} {\bibfnamefont {F.}~\bibnamefont
  {Schwarz}}, \bibinfo {author} {\bibfnamefont {M.}~\bibnamefont {Goldstein}},
  \bibinfo {author} {\bibfnamefont {A.}~\bibnamefont {Dorda}}, \bibinfo
  {author} {\bibfnamefont {E.}~\bibnamefont {Arrigoni}}, \bibinfo {author}
  {\bibfnamefont {A.}~\bibnamefont {Weichselbaum}}, \ and\ \bibinfo {author}
  {\bibfnamefont {J.}~\bibnamefont {von Delft}},\ }\bibfield  {title} {\enquote
  {\bibinfo {title} {Lindblad-driven discretized leads for nonequilibrium
  steady-state transport in quantum impurity models: Recovering the continuum
  limit},}\ }\href {\doibase 10.1103/PhysRevB.94.155142} {\bibfield  {journal}
  {\bibinfo  {journal} {Phys. Rev. B}\ }\textbf {\bibinfo {volume} {94}},\
  \bibinfo {pages} {155142} (\bibinfo {year} {2016})}\BibitemShut {NoStop}%
\bibitem [{\citenamefont {Fugger}\ \emph {et~al.}(2018)\citenamefont {Fugger},
  \citenamefont {Dorda}, \citenamefont {Schwarz}, \citenamefont {von Delft},\
  and\ \citenamefont {Arrigoni}}]{fugger2018}%
  \BibitemOpen
  \bibfield  {author} {\bibinfo {author} {\bibfnamefont {D.~M.}\ \bibnamefont
  {Fugger}}, \bibinfo {author} {\bibfnamefont {A.}~\bibnamefont {Dorda}},
  \bibinfo {author} {\bibfnamefont {F.}~\bibnamefont {Schwarz}}, \bibinfo
  {author} {\bibfnamefont {J.}~\bibnamefont {von Delft}}, \ and\ \bibinfo
  {author} {\bibfnamefont {E.}~\bibnamefont {Arrigoni}},\ }\bibfield  {title}
  {\enquote {\bibinfo {title} {Nonequilibrium kondo effect in a magnetic field:
  auxiliary master equation approach},}\ }\href
  {https://doi.org/10.1088/1367-2630/aa9fdc} {\bibfield  {journal} {\bibinfo
  {journal} {New J. Phys.}\ }\textbf {\bibinfo {volume} {20}},\ \bibinfo
  {pages} {013030} (\bibinfo {year} {2018})}\BibitemShut {NoStop}%
\bibitem [{\citenamefont {Ajisaka}\ \emph {et~al.}(2012)\citenamefont
  {Ajisaka}, \citenamefont {Barra}, \citenamefont {Mej\'{\i}a-Monasterio},\
  and\ \citenamefont {Prosen}}]{ajisaka2012}%
  \BibitemOpen
  \bibfield  {author} {\bibinfo {author} {\bibfnamefont {S.}~\bibnamefont
  {Ajisaka}}, \bibinfo {author} {\bibfnamefont {F.}~\bibnamefont {Barra}},
  \bibinfo {author} {\bibfnamefont {C.}~\bibnamefont {Mej\'{\i}a-Monasterio}},
  \ and\ \bibinfo {author} {\bibfnamefont {T.}~\bibnamefont {Prosen}},\
  }\bibfield  {title} {\enquote {\bibinfo {title} {Nonequlibrium particle and
  energy currents in quantum chains connected to mesoscopic fermi
  reservoirs},}\ }\href {\doibase 10.1103/PhysRevB.86.125111} {\bibfield
  {journal} {\bibinfo  {journal} {Phys. Rev. B}\ }\textbf {\bibinfo {volume}
  {86}},\ \bibinfo {pages} {125111} (\bibinfo {year} {2012})}\BibitemShut
  {NoStop}%
\bibitem [{\citenamefont {Arrigoni}\ \emph {et~al.}(2013)\citenamefont
  {Arrigoni}, \citenamefont {Knap},\ and\ \citenamefont {von~der
  Linden}}]{knap2013}%
  \BibitemOpen
  \bibfield  {author} {\bibinfo {author} {\bibfnamefont {E.}~\bibnamefont
  {Arrigoni}}, \bibinfo {author} {\bibfnamefont {M.}~\bibnamefont {Knap}}, \
  and\ \bibinfo {author} {\bibfnamefont {W.}~\bibnamefont {von~der Linden}},\
  }\bibfield  {title} {\enquote {\bibinfo {title} {Nonequilibrium dynamical
  mean-field theory: An auxiliary quantum master equation approach},}\ }\href
  {\doibase 10.1103/PhysRevLett.110.086403} {\bibfield  {journal} {\bibinfo
  {journal} {Phys. Rev. Lett.}\ }\textbf {\bibinfo {volume} {110}},\ \bibinfo
  {pages} {086403} (\bibinfo {year} {2013})}\BibitemShut {NoStop}%
\bibitem [{\citenamefont {Titvinidze}\ \emph {et~al.}(2015)\citenamefont
  {Titvinidze}, \citenamefont {Dorda}, \citenamefont {von~der Linden},\ and\
  \citenamefont {Arrigoni}}]{titvinidze2015}%
  \BibitemOpen
  \bibfield  {author} {\bibinfo {author} {\bibfnamefont {I.}~\bibnamefont
  {Titvinidze}}, \bibinfo {author} {\bibfnamefont {A.}~\bibnamefont {Dorda}},
  \bibinfo {author} {\bibfnamefont {W.}~\bibnamefont {von~der Linden}}, \ and\
  \bibinfo {author} {\bibfnamefont {E.}~\bibnamefont {Arrigoni}},\ }\bibfield
  {title} {\enquote {\bibinfo {title} {Transport through a correlated
  interface: Auxiliary master equation approach},}\ }\href {\doibase
  10.1103/PhysRevB.92.245125} {\bibfield  {journal} {\bibinfo  {journal} {Phys.
  Rev. B}\ }\textbf {\bibinfo {volume} {92}},\ \bibinfo {pages} {245125}
  (\bibinfo {year} {2015})}\BibitemShut {NoStop}%
\bibitem [{\citenamefont {Titvinidze}\ \emph {et~al.}(2016)\citenamefont
  {Titvinidze}, \citenamefont {Dorda}, \citenamefont {von~der Linden},\ and\
  \citenamefont {Arrigoni}}]{titvinidze2016}%
  \BibitemOpen
  \bibfield  {author} {\bibinfo {author} {\bibfnamefont {I.}~\bibnamefont
  {Titvinidze}}, \bibinfo {author} {\bibfnamefont {A.}~\bibnamefont {Dorda}},
  \bibinfo {author} {\bibfnamefont {W.}~\bibnamefont {von~der Linden}}, \ and\
  \bibinfo {author} {\bibfnamefont {E.}~\bibnamefont {Arrigoni}},\ }\bibfield
  {title} {\enquote {\bibinfo {title} {Resonance effects in correlated
  multilayer heterostructures},}\ }\href {\doibase 10.1103/PhysRevB.94.245142}
  {\bibfield  {journal} {\bibinfo  {journal} {Phys. Rev. B}\ }\textbf {\bibinfo
  {volume} {94}},\ \bibinfo {pages} {245142} (\bibinfo {year}
  {2016})}\BibitemShut {NoStop}%
\bibitem [{\citenamefont {Breuer}\ and\ \citenamefont
  {Petruccione}(2002)}]{breuer2002}%
  \BibitemOpen
  \bibfield  {author} {\bibinfo {author} {\bibfnamefont {H.~P.}\ \bibnamefont
  {Breuer}}\ and\ \bibinfo {author} {\bibfnamefont {F.}~\bibnamefont
  {Petruccione}},\ }\href@noop {} {\emph {\bibinfo {title} {The Theory of Open
  Quantum Systems}}}\ (\bibinfo  {publisher} {Oxford University Press},\
  \bibinfo {year} {2002})\BibitemShut {NoStop}%
\bibitem [{\citenamefont {Carmichael}(2002)}]{carmichael2002}%
  \BibitemOpen
  \bibfield  {author} {\bibinfo {author} {\bibfnamefont {H.~J.}\ \bibnamefont
  {Carmichael}},\ }\href@noop {} {\emph {\bibinfo {title} {Statistical Methods
  in Quantum Optics 1: Master Equations and Fokker-Planck Equations}}}\
  (\bibinfo  {publisher} {Springer},\ \bibinfo {year} {2002})\BibitemShut
  {NoStop}%
\bibitem [{\citenamefont {Hubbard}(1963)}]{hubbard1963}%
  \BibitemOpen
  \bibfield  {author} {\bibinfo {author} {\bibfnamefont {J.}~\bibnamefont
  {Hubbard}},\ }\bibfield  {title} {\enquote {\bibinfo {title} {Electron
  correlations in narrow energy bands},}\ }\href
  {http://www.jstor.org/stable/2414761} {\bibfield  {journal} {\bibinfo
  {journal} {Proceedings of the Royal Society of London. Series A, Mathematical
  and Physical Sciences}\ }\textbf {\bibinfo {volume} {276}},\ \bibinfo {pages}
  {238--257} (\bibinfo {year} {1963})}\BibitemShut {NoStop}%
\bibitem [{\citenamefont {Gutzwiller}(1963)}]{gutzwiller1963}%
  \BibitemOpen
  \bibfield  {author} {\bibinfo {author} {\bibfnamefont {Martin~C.}\
  \bibnamefont {Gutzwiller}},\ }\bibfield  {title} {\enquote {\bibinfo {title}
  {Effect of correlation on the ferromagnetism of transition metals},}\ }\href
  {\doibase 10.1103/PhysRevLett.10.159} {\bibfield  {journal} {\bibinfo
  {journal} {Phys. Rev. Lett.}\ }\textbf {\bibinfo {volume} {10}},\ \bibinfo
  {pages} {159--162} (\bibinfo {year} {1963})}\BibitemShut {NoStop}%
\bibitem [{\citenamefont {Kanamori}(1963)}]{konamori1963}%
  \BibitemOpen
  \bibfield  {author} {\bibinfo {author} {\bibfnamefont {J.}~\bibnamefont
  {Kanamori}},\ }\bibfield  {title} {\enquote {\bibinfo {title} {Electron
  correlation and ferromagnetism of transition metals},}\ }\href {\doibase
  10.1143/PTP.30.275} {\bibfield  {journal} {\bibinfo  {journal} {Progress of
  Theoretical Physics}\ }\textbf {\bibinfo {volume} {30}},\ \bibinfo {pages}
  {275--289} (\bibinfo {year} {1963})}\BibitemShut {NoStop}%
\bibitem [{\citenamefont {Gersch}\ and\ \citenamefont
  {Knollman}(1963)}]{gersch1963}%
  \BibitemOpen
  \bibfield  {author} {\bibinfo {author} {\bibfnamefont {H.~A.}\ \bibnamefont
  {Gersch}}\ and\ \bibinfo {author} {\bibfnamefont {G.~C.}\ \bibnamefont
  {Knollman}},\ }\bibfield  {title} {\enquote {\bibinfo {title} {Quantum cell
  model for bosons},}\ }\href {\doibase 10.1103/PhysRev.129.959} {\bibfield
  {journal} {\bibinfo  {journal} {Phys. Rev.}\ }\textbf {\bibinfo {volume}
  {129}},\ \bibinfo {pages} {959--967} (\bibinfo {year} {1963})}\BibitemShut
  {NoStop}%
\bibitem [{\citenamefont {Jaksch}\ \emph {et~al.}(1998)\citenamefont {Jaksch},
  \citenamefont {Bruder}, \citenamefont {Cirac}, \citenamefont {Gardiner},\
  and\ \citenamefont {Zoller}}]{jaksch1998}%
  \BibitemOpen
  \bibfield  {author} {\bibinfo {author} {\bibfnamefont {D.}~\bibnamefont
  {Jaksch}}, \bibinfo {author} {\bibfnamefont {C.}~\bibnamefont {Bruder}},
  \bibinfo {author} {\bibfnamefont {J.~I.}\ \bibnamefont {Cirac}}, \bibinfo
  {author} {\bibfnamefont {C.~W.}\ \bibnamefont {Gardiner}}, \ and\ \bibinfo
  {author} {\bibfnamefont {P.}~\bibnamefont {Zoller}},\ }\bibfield  {title}
  {\enquote {\bibinfo {title} {Cold bosonic atoms in optical lattices},}\
  }\href {\doibase 10.1103/PhysRevLett.81.3108} {\bibfield  {journal} {\bibinfo
   {journal} {Phys. Rev. Lett.}\ }\textbf {\bibinfo {volume} {81}},\ \bibinfo
  {pages} {3108--3111} (\bibinfo {year} {1998})}\BibitemShut {NoStop}%
\bibitem [{\citenamefont {Greiner}\ \emph {et~al.}(2002)\citenamefont
  {Greiner}, \citenamefont {Mandel}, \citenamefont {Esslinger}, \citenamefont
  {H\"ansch},\ and\ \citenamefont {Bloch}}]{greiner2002}%
  \BibitemOpen
  \bibfield  {author} {\bibinfo {author} {\bibfnamefont {M.}~\bibnamefont
  {Greiner}}, \bibinfo {author} {\bibfnamefont {O.}~\bibnamefont {Mandel}},
  \bibinfo {author} {\bibfnamefont {T.}~\bibnamefont {Esslinger}}, \bibinfo
  {author} {\bibfnamefont {T.~W.}\ \bibnamefont {H\"ansch}}, \ and\ \bibinfo
  {author} {\bibfnamefont {I.}~\bibnamefont {Bloch}},\ }\bibfield  {title}
  {\enquote {\bibinfo {title} {Quantum phase transition from a superfluid to a
  mott insulator in a gas of ultracold atoms},}\ }\href {\doibase
  10.1038/415039a} {\bibfield  {journal} {\bibinfo  {journal} {Nature}\
  }\textbf {\bibinfo {volume} {415}},\ \bibinfo {pages} {39} (\bibinfo {year}
  {2002})}\BibitemShut {NoStop}%
\bibitem [{\citenamefont {J{\"o}rdens}\ \emph {et~al.}(2008)\citenamefont
  {J{\"o}rdens}, \citenamefont {Strohmaier}, \citenamefont {G{\"u}nter},
  \citenamefont {Moritz},\ and\ \citenamefont {Esslinger}}]{joerdens2008}%
  \BibitemOpen
  \bibfield  {author} {\bibinfo {author} {\bibfnamefont {R.}~\bibnamefont
  {J{\"o}rdens}}, \bibinfo {author} {\bibfnamefont {N.}~\bibnamefont
  {Strohmaier}}, \bibinfo {author} {\bibfnamefont {K.}~\bibnamefont
  {G{\"u}nter}}, \bibinfo {author} {\bibfnamefont {H.}~\bibnamefont {Moritz}},
  \ and\ \bibinfo {author} {\bibfnamefont {T.}~\bibnamefont {Esslinger}},\
  }\bibfield  {title} {\enquote {\bibinfo {title} {A mott insulator of
  fermionic atoms in an optical lattice},}\ }\href {\doibase
  10.1038/nature07244} {\bibfield  {journal} {\bibinfo  {journal} {Nature}\
  }\textbf {\bibinfo {volume} {455}},\ \bibinfo {pages} {204} (\bibinfo {year}
  {2008})}\BibitemShut {NoStop}%
\bibitem [{\citenamefont {Schneider}\ \emph {et~al.}(2008)\citenamefont
  {Schneider}, \citenamefont {Hackerm{\"u}ller}, \citenamefont {Will},
  \citenamefont {Best}, \citenamefont {Bloch}, \citenamefont {Costi},
  \citenamefont {Helmes}, \citenamefont {Rasch},\ and\ \citenamefont
  {Rosch}}]{schneider2008}%
  \BibitemOpen
  \bibfield  {author} {\bibinfo {author} {\bibfnamefont {U.}~\bibnamefont
  {Schneider}}, \bibinfo {author} {\bibfnamefont {L.}~\bibnamefont
  {Hackerm{\"u}ller}}, \bibinfo {author} {\bibfnamefont {S.}~\bibnamefont
  {Will}}, \bibinfo {author} {\bibfnamefont {Th.}\ \bibnamefont {Best}},
  \bibinfo {author} {\bibfnamefont {I.}~\bibnamefont {Bloch}}, \bibinfo
  {author} {\bibfnamefont {T.~A.}\ \bibnamefont {Costi}}, \bibinfo {author}
  {\bibfnamefont {R.~W.}\ \bibnamefont {Helmes}}, \bibinfo {author}
  {\bibfnamefont {D.}~\bibnamefont {Rasch}}, \ and\ \bibinfo {author}
  {\bibfnamefont {A.}~\bibnamefont {Rosch}},\ }\bibfield  {title} {\enquote
  {\bibinfo {title} {Metallic and insulating phases of repulsively interacting
  fermions in a 3d optical lattice},}\ }\href {\doibase
  10.1126/science.1165449} {\bibfield  {journal} {\bibinfo  {journal}
  {Science}\ }\textbf {\bibinfo {volume} {322}},\ \bibinfo {pages} {1520--1525}
  (\bibinfo {year} {2008})},\ \Eprint
  {http://arxiv.org/abs/http://science.sciencemag.org/content/322/5907/1520.full.pdf}
  {http://science.sciencemag.org/content/322/5907/1520.full.pdf} \BibitemShut
  {NoStop}%
\bibitem [{\citenamefont {Varney}\ \emph {et~al.}(2009)\citenamefont {Varney},
  \citenamefont {Lee}, \citenamefont {Bai}, \citenamefont {Chiesa},
  \citenamefont {Jarrell},\ and\ \citenamefont {Scalettar}}]{varney2009}%
  \BibitemOpen
  \bibfield  {author} {\bibinfo {author} {\bibfnamefont {C.~N.}\ \bibnamefont
  {Varney}}, \bibinfo {author} {\bibfnamefont {C.-R.}\ \bibnamefont {Lee}},
  \bibinfo {author} {\bibfnamefont {Z.~J.}\ \bibnamefont {Bai}}, \bibinfo
  {author} {\bibfnamefont {S.}~\bibnamefont {Chiesa}}, \bibinfo {author}
  {\bibfnamefont {M.}~\bibnamefont {Jarrell}}, \ and\ \bibinfo {author}
  {\bibfnamefont {R.~T.}\ \bibnamefont {Scalettar}},\ }\bibfield  {title}
  {\enquote {\bibinfo {title} {Quantum monte carlo study of the two-dimensional
  fermion hubbard model},}\ }\href {\doibase 10.1103/PhysRevB.80.075116}
  {\bibfield  {journal} {\bibinfo  {journal} {Phys. Rev. B}\ }\textbf {\bibinfo
  {volume} {80}},\ \bibinfo {pages} {075116} (\bibinfo {year}
  {2009})}\BibitemShut {NoStop}%
\bibitem [{\citenamefont {Micnas}\ \emph {et~al.}(1988)\citenamefont {Micnas},
  \citenamefont {Ranninger}, \citenamefont {Robaszkiewicz},\ and\ \citenamefont
  {Tabor}}]{micnas1988}%
  \BibitemOpen
  \bibfield  {author} {\bibinfo {author} {\bibfnamefont {R.}~\bibnamefont
  {Micnas}}, \bibinfo {author} {\bibfnamefont {J.}~\bibnamefont {Ranninger}},
  \bibinfo {author} {\bibfnamefont {S.}~\bibnamefont {Robaszkiewicz}}, \ and\
  \bibinfo {author} {\bibfnamefont {S.}~\bibnamefont {Tabor}},\ }\bibfield
  {title} {\enquote {\bibinfo {title} {Superconductivity in a narrow-band
  system with intersite electron pairing in two dimensions: A mean-field
  study},}\ }\href {\doibase 10.1103/PhysRevB.37.9410} {\bibfield  {journal}
  {\bibinfo  {journal} {Phys. Rev. B}\ }\textbf {\bibinfo {volume} {37}},\
  \bibinfo {pages} {9410--9422} (\bibinfo {year} {1988})}\BibitemShut {NoStop}%
\bibitem [{\citenamefont {Dagotto}\ \emph {et~al.}(1994)\citenamefont
  {Dagotto}, \citenamefont {Riera}, \citenamefont {Chen}, \citenamefont
  {Moreo}, \citenamefont {Nazarenko}, \citenamefont {Alcaraz},\ and\
  \citenamefont {Ortolani}}]{dagotto1994}%
  \BibitemOpen
  \bibfield  {author} {\bibinfo {author} {\bibfnamefont {E.}~\bibnamefont
  {Dagotto}}, \bibinfo {author} {\bibfnamefont {J.}~\bibnamefont {Riera}},
  \bibinfo {author} {\bibfnamefont {Y.~C.}\ \bibnamefont {Chen}}, \bibinfo
  {author} {\bibfnamefont {A.}~\bibnamefont {Moreo}}, \bibinfo {author}
  {\bibfnamefont {A.}~\bibnamefont {Nazarenko}}, \bibinfo {author}
  {\bibfnamefont {F.}~\bibnamefont {Alcaraz}}, \ and\ \bibinfo {author}
  {\bibfnamefont {F.}~\bibnamefont {Ortolani}},\ }\bibfield  {title} {\enquote
  {\bibinfo {title} {Superconductivity near phase separation in models of
  correlated electrons},}\ }\href {\doibase 10.1103/PhysRevB.49.3548}
  {\bibfield  {journal} {\bibinfo  {journal} {Phys. Rev. B}\ }\textbf {\bibinfo
  {volume} {49}},\ \bibinfo {pages} {3548--3565} (\bibinfo {year}
  {1994})}\BibitemShut {NoStop}%
\bibitem [{\citenamefont {Chattopadhyay}\ and\ \citenamefont
  {Gaitonde}(1997)}]{chattopadhyay1997}%
  \BibitemOpen
  \bibfield  {author} {\bibinfo {author} {\bibfnamefont {B.}~\bibnamefont
  {Chattopadhyay}}\ and\ \bibinfo {author} {\bibfnamefont {D.~M.}\ \bibnamefont
  {Gaitonde}},\ }\bibfield  {title} {\enquote {\bibinfo {title} {Phase diagram
  of the half-filled extended hubbard model in two dimensions},}\ }\href
  {\doibase 10.1103/PhysRevB.55.15364} {\bibfield  {journal} {\bibinfo
  {journal} {Phys. Rev. B}\ }\textbf {\bibinfo {volume} {55}},\ \bibinfo
  {pages} {15364--15367} (\bibinfo {year} {1997})}\BibitemShut {NoStop}%
\bibitem [{\citenamefont {Aichhorn}\ \emph {et~al.}(2004)\citenamefont
  {Aichhorn}, \citenamefont {Evertz}, \citenamefont {von~der Linden},\ and\
  \citenamefont {Potthoff}}]{aichhorn2004}%
  \BibitemOpen
  \bibfield  {author} {\bibinfo {author} {\bibfnamefont {M.}~\bibnamefont
  {Aichhorn}}, \bibinfo {author} {\bibfnamefont {H.~G.}\ \bibnamefont
  {Evertz}}, \bibinfo {author} {\bibfnamefont {W.}~\bibnamefont {von~der
  Linden}}, \ and\ \bibinfo {author} {\bibfnamefont {M.}~\bibnamefont
  {Potthoff}},\ }\bibfield  {title} {\enquote {\bibinfo {title} {Charge
  ordering in extended hubbard models: Variational cluster approach},}\ }\href
  {\doibase 10.1103/PhysRevB.70.235107} {\bibfield  {journal} {\bibinfo
  {journal} {Phys. Rev. B}\ }\textbf {\bibinfo {volume} {70}},\ \bibinfo
  {pages} {235107} (\bibinfo {year} {2004})}\BibitemShut {NoStop}%
\bibitem [{\citenamefont {Kapcia}\ \emph {et~al.}(2017)\citenamefont {Kapcia},
  \citenamefont {Robaszkiewicz}, \citenamefont {Capone},\ and\ \citenamefont
  {Amaricci}}]{kapcia2017}%
  \BibitemOpen
  \bibfield  {author} {\bibinfo {author} {\bibfnamefont {K.~J.}\ \bibnamefont
  {Kapcia}}, \bibinfo {author} {\bibfnamefont {S.}~\bibnamefont
  {Robaszkiewicz}}, \bibinfo {author} {\bibfnamefont {M.}~\bibnamefont
  {Capone}}, \ and\ \bibinfo {author} {\bibfnamefont {A.}~\bibnamefont
  {Amaricci}},\ }\bibfield  {title} {\enquote {\bibinfo {title} {Doping-driven
  metal-insulator transitions and charge orderings in the extended hubbard
  model},}\ }\href {\doibase 10.1103/PhysRevB.95.125112} {\bibfield  {journal}
  {\bibinfo  {journal} {Phys. Rev. B}\ }\textbf {\bibinfo {volume} {95}},\
  \bibinfo {pages} {125112} (\bibinfo {year} {2017})}\BibitemShut {NoStop}%
\bibitem [{\citenamefont {Terletska}\ \emph {et~al.}(2017)\citenamefont
  {Terletska}, \citenamefont {Chen},\ and\ \citenamefont
  {Gull}}]{terletska2017}%
  \BibitemOpen
  \bibfield  {author} {\bibinfo {author} {\bibfnamefont {H.}~\bibnamefont
  {Terletska}}, \bibinfo {author} {\bibfnamefont {T.}~\bibnamefont {Chen}}, \
  and\ \bibinfo {author} {\bibfnamefont {E.}~\bibnamefont {Gull}},\ }\bibfield
  {title} {\enquote {\bibinfo {title} {Charge ordering and correlation effects
  in the extended hubbard model},}\ }\href {\doibase
  10.1103/PhysRevB.95.115149} {\bibfield  {journal} {\bibinfo  {journal} {Phys.
  Rev. B}\ }\textbf {\bibinfo {volume} {95}},\ \bibinfo {pages} {115149}
  (\bibinfo {year} {2017})}\BibitemShut {NoStop}%
\bibitem [{\citenamefont {Terletska}\ \emph {et~al.}(2018)\citenamefont
  {Terletska}, \citenamefont {Chen}, \citenamefont {Paki},\ and\ \citenamefont
  {Gull}}]{terletska2018}%
  \BibitemOpen
  \bibfield  {author} {\bibinfo {author} {\bibfnamefont {H.}~\bibnamefont
  {Terletska}}, \bibinfo {author} {\bibfnamefont {T.}~\bibnamefont {Chen}},
  \bibinfo {author} {\bibfnamefont {J.}~\bibnamefont {Paki}}, \ and\ \bibinfo
  {author} {\bibfnamefont {E.}~\bibnamefont {Gull}},\ }\bibfield  {title}
  {\enquote {\bibinfo {title} {Charge ordering and nonlocal correlations in the
  doped extended hubbard model},}\ }\href {\doibase 10.1103/PhysRevB.97.115117}
  {\bibfield  {journal} {\bibinfo  {journal} {Phys. Rev. B}\ }\textbf {\bibinfo
  {volume} {97}},\ \bibinfo {pages} {115117} (\bibinfo {year}
  {2018})}\BibitemShut {NoStop}%
\bibitem [{\citenamefont {McKenzie}\ \emph {et~al.}(2001)\citenamefont
  {McKenzie}, \citenamefont {Merino}, \citenamefont {Marston},\ and\
  \citenamefont {Sushkov}}]{mckenzie2001}%
  \BibitemOpen
  \bibfield  {author} {\bibinfo {author} {\bibfnamefont {Ross~H.}\ \bibnamefont
  {McKenzie}}, \bibinfo {author} {\bibfnamefont {J.}~\bibnamefont {Merino}},
  \bibinfo {author} {\bibfnamefont {J.~B.}\ \bibnamefont {Marston}}, \ and\
  \bibinfo {author} {\bibfnamefont {O.~P.}\ \bibnamefont {Sushkov}},\
  }\bibfield  {title} {\enquote {\bibinfo {title} {Charge ordering and
  antiferromagnetic exchange in layered molecular crystals of the
  $\ensuremath{\theta}$ type},}\ }\href {\doibase 10.1103/PhysRevB.64.085109}
  {\bibfield  {journal} {\bibinfo  {journal} {Phys. Rev. B}\ }\textbf {\bibinfo
  {volume} {64}},\ \bibinfo {pages} {085109} (\bibinfo {year}
  {2001})}\BibitemShut {NoStop}%
\bibitem [{\citenamefont {Merino}\ and\ \citenamefont
  {McKenzie}(2001)}]{merino2001}%
  \BibitemOpen
  \bibfield  {author} {\bibinfo {author} {\bibfnamefont {Jaime}\ \bibnamefont
  {Merino}}\ and\ \bibinfo {author} {\bibfnamefont {Ross~H.}\ \bibnamefont
  {McKenzie}},\ }\bibfield  {title} {\enquote {\bibinfo {title}
  {Superconductivity mediated by charge fluctuations in layered molecular
  crystals},}\ }\href {\doibase 10.1103/PhysRevLett.87.237002} {\bibfield
  {journal} {\bibinfo  {journal} {Phys. Rev. Lett.}\ }\textbf {\bibinfo
  {volume} {87}},\ \bibinfo {pages} {237002} (\bibinfo {year}
  {2001})}\BibitemShut {NoStop}%
\bibitem [{\citenamefont {Kobayashi}\ \emph {et~al.}(2004)\citenamefont
  {Kobayashi}, \citenamefont {Tanaka}, \citenamefont {Ogata},\ and\
  \citenamefont {Suzumura}}]{kobayashi2004}%
  \BibitemOpen
  \bibfield  {author} {\bibinfo {author} {\bibfnamefont {A.}~\bibnamefont
  {Kobayashi}}, \bibinfo {author} {\bibfnamefont {Y.}~\bibnamefont {Tanaka}},
  \bibinfo {author} {\bibfnamefont {M.}~\bibnamefont {Ogata}}, \ and\ \bibinfo
  {author} {\bibfnamefont {Y.}~\bibnamefont {Suzumura}},\ }\bibfield  {title}
  {\enquote {\bibinfo {title} {Charge-fluctuation-induced superconducting state
  in two-dimensional quarter-filled electron systems},}\ }\href {\doibase
  10.1143/JPSJ.73.1115} {\bibfield  {journal} {\bibinfo  {journal} {Journal of
  the Physical Society of Japan}\ }\textbf {\bibinfo {volume} {73}},\ \bibinfo
  {pages} {1115--1118} (\bibinfo {year} {2004})}\BibitemShut {NoStop}%
\bibitem [{\citenamefont {Calandra}\ \emph {et~al.}(2002)\citenamefont
  {Calandra}, \citenamefont {Merino},\ and\ \citenamefont
  {McKenzie}}]{calandra2002}%
  \BibitemOpen
  \bibfield  {author} {\bibinfo {author} {\bibfnamefont {M.}~\bibnamefont
  {Calandra}}, \bibinfo {author} {\bibfnamefont {J.}~\bibnamefont {Merino}}, \
  and\ \bibinfo {author} {\bibfnamefont {Ross~H.}\ \bibnamefont {McKenzie}},\
  }\bibfield  {title} {\enquote {\bibinfo {title} {Metal-insulator transition
  and charge ordering in the extended hubbard model at one-quarter filling},}\
  }\href {\doibase 10.1103/PhysRevB.66.195102} {\bibfield  {journal} {\bibinfo
  {journal} {Phys. Rev. B}\ }\textbf {\bibinfo {volume} {66}},\ \bibinfo
  {pages} {195102} (\bibinfo {year} {2002})}\BibitemShut {NoStop}%
\bibitem [{\citenamefont {Garg}\ \emph {et~al.}(2014)\citenamefont {Garg},
  \citenamefont {Krishnamurthy},\ and\ \citenamefont {Randeria}}]{garg2014}%
  \BibitemOpen
  \bibfield  {author} {\bibinfo {author} {\bibfnamefont {A.}~\bibnamefont
  {Garg}}, \bibinfo {author} {\bibfnamefont {H.~R.}\ \bibnamefont
  {Krishnamurthy}}, \ and\ \bibinfo {author} {\bibfnamefont {M.}~\bibnamefont
  {Randeria}},\ }\bibfield  {title} {\enquote {\bibinfo {title} {Doping a
  correlated band insulator: A new route to half-metallic behavior},}\ }\href
  {\doibase 10.1103/PhysRevLett.112.106406} {\bibfield  {journal} {\bibinfo
  {journal} {Phys. Rev. Lett.}\ }\textbf {\bibinfo {volume} {112}},\ \bibinfo
  {pages} {106406} (\bibinfo {year} {2014})}\BibitemShut {NoStop}%
\bibitem [{\citenamefont {Baranov}\ \emph {et~al.}(2012)\citenamefont
  {Baranov}, \citenamefont {Dalmonte}, \citenamefont {Pupillo},\ and\
  \citenamefont {Zoller}}]{baranov2012}%
  \BibitemOpen
  \bibfield  {author} {\bibinfo {author} {\bibfnamefont {M.~A.}\ \bibnamefont
  {Baranov}}, \bibinfo {author} {\bibfnamefont {M.}~\bibnamefont {Dalmonte}},
  \bibinfo {author} {\bibfnamefont {G.}~\bibnamefont {Pupillo}}, \ and\
  \bibinfo {author} {\bibfnamefont {P.}~\bibnamefont {Zoller}},\ }\bibfield
  {title} {\enquote {\bibinfo {title} {Condensed matter theory of dipolar
  quantum gases},}\ }\href {\doibase 10.1021/cr2003568} {\bibfield  {journal}
  {\bibinfo  {journal} {Chemical Reviews}\ }\textbf {\bibinfo {volume} {112}},\
  \bibinfo {pages} {5012--5061} (\bibinfo {year} {2012})}\BibitemShut {NoStop}%
\bibitem [{\citenamefont {Yan}\ \emph {et~al.}(2013)\citenamefont {Yan},
  \citenamefont {Moses}, \citenamefont {Gadway}, \citenamefont {Covey},
  \citenamefont {Hazzard}, \citenamefont {Rey}, \citenamefont {Jin},\ and\
  \citenamefont {Ye}}]{yan2013}%
  \BibitemOpen
  \bibfield  {author} {\bibinfo {author} {\bibfnamefont {B.}~\bibnamefont
  {Yan}}, \bibinfo {author} {\bibfnamefont {S.~A.}\ \bibnamefont {Moses}},
  \bibinfo {author} {\bibfnamefont {B.}~\bibnamefont {Gadway}}, \bibinfo
  {author} {\bibfnamefont {J.~P.}\ \bibnamefont {Covey}}, \bibinfo {author}
  {\bibfnamefont {K.~R.~A.}\ \bibnamefont {Hazzard}}, \bibinfo {author}
  {\bibfnamefont {A.~M.}\ \bibnamefont {Rey}}, \bibinfo {author} {\bibfnamefont
  {D.~S.}\ \bibnamefont {Jin}}, \ and\ \bibinfo {author} {\bibfnamefont
  {J.}~\bibnamefont {Ye}},\ }\bibfield  {title} {\enquote {\bibinfo {title}
  {Observation of dipolar spin-exchange interactions with lattice-confined
  polar molecules},}\ }\href {\doibase 10.1038/nature12483} {\bibfield
  {journal} {\bibinfo  {journal} {Nature}\ }\textbf {\bibinfo {volume} {501}},\
  \bibinfo {pages} {521} (\bibinfo {year} {2013})}\BibitemShut {NoStop}%
\bibitem [{\citenamefont {Baier}\ \emph {et~al.}(2016)\citenamefont {Baier},
  \citenamefont {Mark}, \citenamefont {Petter}, \citenamefont {Aikawa},
  \citenamefont {Chomaz}, \citenamefont {Cai}, \citenamefont {Baranov},
  \citenamefont {Zoller},\ and\ \citenamefont {Ferlaino}}]{baier2016}%
  \BibitemOpen
  \bibfield  {author} {\bibinfo {author} {\bibfnamefont {S.}~\bibnamefont
  {Baier}}, \bibinfo {author} {\bibfnamefont {M.~J.}\ \bibnamefont {Mark}},
  \bibinfo {author} {\bibfnamefont {D.}~\bibnamefont {Petter}}, \bibinfo
  {author} {\bibfnamefont {K.}~\bibnamefont {Aikawa}}, \bibinfo {author}
  {\bibfnamefont {L.}~\bibnamefont {Chomaz}}, \bibinfo {author} {\bibfnamefont
  {Z.}~\bibnamefont {Cai}}, \bibinfo {author} {\bibfnamefont {M.}~\bibnamefont
  {Baranov}}, \bibinfo {author} {\bibfnamefont {P.}~\bibnamefont {Zoller}}, \
  and\ \bibinfo {author} {\bibfnamefont {F.}~\bibnamefont {Ferlaino}},\
  }\bibfield  {title} {\enquote {\bibinfo {title} {Extended bose-hubbard models
  with ultracold magnetic atoms},}\ }\href {\doibase 10.1126/science.aac9812}
  {\bibfield  {journal} {\bibinfo  {journal} {Science}\ }\textbf {\bibinfo
  {volume} {352}},\ \bibinfo {pages} {201--205} (\bibinfo {year} {2016})},\
  \Eprint
  {http://arxiv.org/abs/http://science.sciencemag.org/content/352/6282/201.full.pdf}
  {http://science.sciencemag.org/content/352/6282/201.full.pdf} \BibitemShut
  {NoStop}%
\bibitem [{\citenamefont {Gallagher}(1988)}]{gallagher1988}%
  \BibitemOpen
  \bibfield  {author} {\bibinfo {author} {\bibfnamefont {T.~F.}\ \bibnamefont
  {Gallagher}},\ }\bibfield  {title} {\enquote {\bibinfo {title} {Rydberg
  atoms},}\ }\href {https://doi.org/10.1088/0034-4885/51/2/001} {\bibfield
  {journal} {\bibinfo  {journal} {Reports on Progress in Physics}\ }\textbf
  {\bibinfo {volume} {51}},\ \bibinfo {pages} {143} (\bibinfo {year}
  {1988})}\BibitemShut {NoStop}%
\bibitem [{\citenamefont {Schau{\ss}}\ \emph {et~al.}(2012)\citenamefont
  {Schau{\ss}}, \citenamefont {Cheneau}, \citenamefont {Endres}, \citenamefont
  {Fukuhara}, \citenamefont {Hild}, \citenamefont {Omran}, \citenamefont
  {Pohl}, \citenamefont {Gross}, \citenamefont {Kuhr},\ and\ \citenamefont
  {Bloch}}]{schauss2012}%
  \BibitemOpen
  \bibfield  {author} {\bibinfo {author} {\bibfnamefont {P.}~\bibnamefont
  {Schau{\ss}}}, \bibinfo {author} {\bibfnamefont {M.}~\bibnamefont {Cheneau}},
  \bibinfo {author} {\bibfnamefont {M.}~\bibnamefont {Endres}}, \bibinfo
  {author} {\bibfnamefont {T.}~\bibnamefont {Fukuhara}}, \bibinfo {author}
  {\bibfnamefont {S.}~\bibnamefont {Hild}}, \bibinfo {author} {\bibfnamefont
  {A.}~\bibnamefont {Omran}}, \bibinfo {author} {\bibfnamefont
  {T.}~\bibnamefont {Pohl}}, \bibinfo {author} {\bibfnamefont {C.}~\bibnamefont
  {Gross}}, \bibinfo {author} {\bibfnamefont {S.}~\bibnamefont {Kuhr}}, \ and\
  \bibinfo {author} {\bibfnamefont {I.}~\bibnamefont {Bloch}},\ }\bibfield
  {title} {\enquote {\bibinfo {title} {Observation of spatially ordered
  structures in a two-dimensional rydberg gas},}\ }\href
  {http://dx.doi.org/10.1038/nature11596} {\bibfield  {journal} {\bibinfo
  {journal} {Nature}\ }\textbf {\bibinfo {volume} {491}},\ \bibinfo {pages}
  {87} (\bibinfo {year} {2012})}\BibitemShut {NoStop}%
\bibitem [{\citenamefont {Schau{\ss}}\ \emph {et~al.}(2015)\citenamefont
  {Schau{\ss}}, \citenamefont {Zeiher}, \citenamefont {Fukuhara}, \citenamefont
  {Hild}, \citenamefont {Cheneau}, \citenamefont {Macr{\`\i}}, \citenamefont
  {Pohl}, \citenamefont {Bloch},\ and\ \citenamefont {Gross}}]{schauss2015}%
  \BibitemOpen
  \bibfield  {author} {\bibinfo {author} {\bibfnamefont {P.}~\bibnamefont
  {Schau{\ss}}}, \bibinfo {author} {\bibfnamefont {J.}~\bibnamefont {Zeiher}},
  \bibinfo {author} {\bibfnamefont {T.}~\bibnamefont {Fukuhara}}, \bibinfo
  {author} {\bibfnamefont {S.}~\bibnamefont {Hild}}, \bibinfo {author}
  {\bibfnamefont {M.}~\bibnamefont {Cheneau}}, \bibinfo {author} {\bibfnamefont
  {T.}~\bibnamefont {Macr{\`\i}}}, \bibinfo {author} {\bibfnamefont
  {T.}~\bibnamefont {Pohl}}, \bibinfo {author} {\bibfnamefont {I.}~\bibnamefont
  {Bloch}}, \ and\ \bibinfo {author} {\bibfnamefont {C.}~\bibnamefont
  {Gross}},\ }\bibfield  {title} {\enquote {\bibinfo {title} {Crystallization
  in ising quantum magnets},}\ }\href {\doibase 10.1126/science.1258351}
  {\bibfield  {journal} {\bibinfo  {journal} {Science}\ }\textbf {\bibinfo
  {volume} {347}},\ \bibinfo {pages} {1455--1458} (\bibinfo {year}
  {2015})}\BibitemShut {NoStop}%
\bibitem [{\citenamefont {Zeiher}\ \emph {et~al.}(2017)\citenamefont {Zeiher},
  \citenamefont {Choi}, \citenamefont {Rubio-Abadal}, \citenamefont {Pohl},
  \citenamefont {van Bijnen}, \citenamefont {Bloch},\ and\ \citenamefont
  {Gross}}]{zeiher2017}%
  \BibitemOpen
  \bibfield  {author} {\bibinfo {author} {\bibfnamefont {J.}~\bibnamefont
  {Zeiher}}, \bibinfo {author} {\bibfnamefont {J.}~\bibnamefont {Choi}},
  \bibinfo {author} {\bibfnamefont {A.}~\bibnamefont {Rubio-Abadal}}, \bibinfo
  {author} {\bibfnamefont {T.}~\bibnamefont {Pohl}}, \bibinfo {author}
  {\bibfnamefont {R.}~\bibnamefont {van Bijnen}}, \bibinfo {author}
  {\bibfnamefont {I.}~\bibnamefont {Bloch}}, \ and\ \bibinfo {author}
  {\bibfnamefont {C.}~\bibnamefont {Gross}},\ }\bibfield  {title} {\enquote
  {\bibinfo {title} {Coherent many-body spin dynamics in a long-range
  interacting ising chain},}\ }\href {\doibase 10.1103/PhysRevX.7.041063}
  {\bibfield  {journal} {\bibinfo  {journal} {Phys. Rev. X}\ }\textbf {\bibinfo
  {volume} {7}},\ \bibinfo {pages} {041063} (\bibinfo {year}
  {2017})}\BibitemShut {NoStop}%
\bibitem [{\citenamefont {Schauss}(2018)}]{schauss2018}%
  \BibitemOpen
  \bibfield  {author} {\bibinfo {author} {\bibfnamefont {P.}~\bibnamefont
  {Schauss}},\ }\bibfield  {title} {\enquote {\bibinfo {title} {Quantum
  simulation of transverse ising models with rydberg atoms},}\ }\href
  {https://doi.org/10.1088/2058-9565/aa9c59} {\bibfield  {journal} {\bibinfo
  {journal} {Quantum Science and Technology}\ }\textbf {\bibinfo {volume}
  {3}},\ \bibinfo {pages} {023001} (\bibinfo {year} {2018})}\BibitemShut
  {NoStop}%
\bibitem [{\citenamefont {Pohl}\ \emph {et~al.}(2010)\citenamefont {Pohl},
  \citenamefont {Demler},\ and\ \citenamefont {Lukin}}]{pohl2010}%
  \BibitemOpen
  \bibfield  {author} {\bibinfo {author} {\bibfnamefont {T.}~\bibnamefont
  {Pohl}}, \bibinfo {author} {\bibfnamefont {E.}~\bibnamefont {Demler}}, \ and\
  \bibinfo {author} {\bibfnamefont {M.~D.}\ \bibnamefont {Lukin}},\ }\bibfield
  {title} {\enquote {\bibinfo {title} {Dynamical crystallization in the dipole
  blockade of ultracold atoms},}\ }\href {\doibase
  10.1103/PhysRevLett.104.043002} {\bibfield  {journal} {\bibinfo  {journal}
  {Phys. Rev. Lett.}\ }\textbf {\bibinfo {volume} {104}},\ \bibinfo {pages}
  {043002} (\bibinfo {year} {2010})}\BibitemShut {NoStop}%
\bibitem [{\citenamefont {Schachenmayer}\ \emph {et~al.}(2010)\citenamefont
  {Schachenmayer}, \citenamefont {Lesanovsky}, \citenamefont {Micheli},\ and\
  \citenamefont {Daley}}]{schachenmayer2010}%
  \BibitemOpen
  \bibfield  {author} {\bibinfo {author} {\bibfnamefont {J.}~\bibnamefont
  {Schachenmayer}}, \bibinfo {author} {\bibfnamefont {I.}~\bibnamefont
  {Lesanovsky}}, \bibinfo {author} {\bibfnamefont {A.}~\bibnamefont {Micheli}},
  \ and\ \bibinfo {author} {\bibfnamefont {A.~J.}\ \bibnamefont {Daley}},\
  }\bibfield  {title} {\enquote {\bibinfo {title} {Dynamical crystal creation
  with polar molecules or rydberg atoms in optical lattices},}\ }\href
  {http://stacks.iop.org/1367-2630/12/i=10/a=103044} {\bibfield  {journal}
  {\bibinfo  {journal} {New Journal of Physics}\ }\textbf {\bibinfo {volume}
  {12}},\ \bibinfo {pages} {103044} (\bibinfo {year} {2010})}\BibitemShut
  {NoStop}%
\bibitem [{\citenamefont {Weimer}\ and\ \citenamefont
  {B\"uchler}(2010)}]{weimer2010}%
  \BibitemOpen
  \bibfield  {author} {\bibinfo {author} {\bibfnamefont {H.}~\bibnamefont
  {Weimer}}\ and\ \bibinfo {author} {\bibfnamefont {H.~P.}\ \bibnamefont
  {B\"uchler}},\ }\bibfield  {title} {\enquote {\bibinfo {title} {Two-stage
  melting in systems of strongly interacting rydberg atoms},}\ }\href {\doibase
  10.1103/PhysRevLett.105.230403} {\bibfield  {journal} {\bibinfo  {journal}
  {Phys. Rev. Lett.}\ }\textbf {\bibinfo {volume} {105}},\ \bibinfo {pages}
  {230403} (\bibinfo {year} {2010})}\BibitemShut {NoStop}%
\bibitem [{\citenamefont {Vermersch}\ \emph {et~al.}(2015)\citenamefont
  {Vermersch}, \citenamefont {Punk}, \citenamefont {Glaetzle}, \citenamefont
  {Gross},\ and\ \citenamefont {Zoller}}]{vermersch2015}%
  \BibitemOpen
  \bibfield  {author} {\bibinfo {author} {\bibfnamefont {B.}~\bibnamefont
  {Vermersch}}, \bibinfo {author} {\bibfnamefont {M.}~\bibnamefont {Punk}},
  \bibinfo {author} {\bibfnamefont {A.~W.}\ \bibnamefont {Glaetzle}}, \bibinfo
  {author} {\bibfnamefont {C.}~\bibnamefont {Gross}}, \ and\ \bibinfo {author}
  {\bibfnamefont {P.}~\bibnamefont {Zoller}},\ }\bibfield  {title} {\enquote
  {\bibinfo {title} {Dynamical preparation of laser-excited anisotropic rydberg
  crystals in 2d optical lattices},}\ }\href
  {http://stacks.iop.org/1367-2630/17/i=1/a=013008} {\bibfield  {journal}
  {\bibinfo  {journal} {New Journal of Physics}\ }\textbf {\bibinfo {volume}
  {17}},\ \bibinfo {pages} {013008} (\bibinfo {year} {2015})}\BibitemShut
  {NoStop}%
\bibitem [{\citenamefont {Lauer}\ \emph {et~al.}(2012)\citenamefont {Lauer},
  \citenamefont {Muth},\ and\ \citenamefont {Fleischhauer}}]{lauer2012}%
  \BibitemOpen
  \bibfield  {author} {\bibinfo {author} {\bibfnamefont {Achim}\ \bibnamefont
  {Lauer}}, \bibinfo {author} {\bibfnamefont {Dominik}\ \bibnamefont {Muth}}, \
  and\ \bibinfo {author} {\bibfnamefont {Michael}\ \bibnamefont
  {Fleischhauer}},\ }\bibfield  {title} {\enquote {\bibinfo {title}
  {Transport-induced melting of crystals of rydberg dressed atoms in a
  one-dimensional lattice},}\ }\href
  {http://stacks.iop.org/1367-2630/14/i=9/a=095009} {\bibfield  {journal}
  {\bibinfo  {journal} {New Journal of Physics}\ }\textbf {\bibinfo {volume}
  {14}},\ \bibinfo {pages} {095009} (\bibinfo {year} {2012})}\BibitemShut
  {NoStop}%
\bibitem [{\citenamefont {Gei\ss{}ler}\ \emph {et~al.}(2017)\citenamefont
  {Gei\ss{}ler}, \citenamefont {Vasi\ifmmode~\acute{c}\else \'{c}\fi{}},\ and\
  \citenamefont {Hofstetter}}]{geissler2017}%
  \BibitemOpen
  \bibfield  {author} {\bibinfo {author} {\bibfnamefont {A.}~\bibnamefont
  {Gei\ss{}ler}}, \bibinfo {author} {\bibfnamefont {I.}~\bibnamefont
  {Vasi\ifmmode~\acute{c}\else \'{c}\fi{}}}, \ and\ \bibinfo {author}
  {\bibfnamefont {W.}~\bibnamefont {Hofstetter}},\ }\bibfield  {title}
  {\enquote {\bibinfo {title} {Condensation versus long-range interaction:
  Competing quantum phases in bosonic optical lattice systems at near-resonant
  rydberg dressing},}\ }\href {\doibase 10.1103/PhysRevA.95.063608} {\bibfield
  {journal} {\bibinfo  {journal} {Phys. Rev. A}\ }\textbf {\bibinfo {volume}
  {95}},\ \bibinfo {pages} {063608} (\bibinfo {year} {2017})}\BibitemShut
  {NoStop}%
\bibitem [{\citenamefont {Li}\ \emph {et~al.}(2018)\citenamefont {Li},
  \citenamefont {Gei\ss{}ler}, \citenamefont {Hofstetter},\ and\ \citenamefont
  {Li}}]{li2018}%
  \BibitemOpen
  \bibfield  {author} {\bibinfo {author} {\bibfnamefont {Y.}~\bibnamefont
  {Li}}, \bibinfo {author} {\bibfnamefont {A.}~\bibnamefont {Gei\ss{}ler}},
  \bibinfo {author} {\bibfnamefont {W.}~\bibnamefont {Hofstetter}}, \ and\
  \bibinfo {author} {\bibfnamefont {W.}~\bibnamefont {Li}},\ }\bibfield
  {title} {\enquote {\bibinfo {title} {Supersolidity of lattice bosons immersed
  in strongly correlated rydberg dressed atoms},}\ }\href {\doibase
  10.1103/PhysRevA.97.023619} {\bibfield  {journal} {\bibinfo  {journal} {Phys.
  Rev. A}\ }\textbf {\bibinfo {volume} {97}},\ \bibinfo {pages} {023619}
  (\bibinfo {year} {2018})}\BibitemShut {NoStop}%
\bibitem [{\citenamefont {Weimer}(2015)}]{weimer2015}%
  \BibitemOpen
  \bibfield  {author} {\bibinfo {author} {\bibfnamefont {H.}~\bibnamefont
  {Weimer}},\ }\bibfield  {title} {\enquote {\bibinfo {title} {Variational
  principle for steady states of dissipative quantum many-body systems},}\
  }\href {\doibase 10.1103/PhysRevLett.114.040402} {\bibfield  {journal}
  {\bibinfo  {journal} {Phys. Rev. Lett.}\ }\textbf {\bibinfo {volume} {114}},\
  \bibinfo {pages} {040402} (\bibinfo {year} {2015})}\BibitemShut {NoStop}%
\bibitem [{\citenamefont {Lichtenstein}\ and\ \citenamefont
  {Katsnelson}(2000)}]{lichtenstein2000}%
  \BibitemOpen
  \bibfield  {author} {\bibinfo {author} {\bibfnamefont {A.~I.}\ \bibnamefont
  {Lichtenstein}}\ and\ \bibinfo {author} {\bibfnamefont {M.~I.}\ \bibnamefont
  {Katsnelson}},\ }\bibfield  {title} {\enquote {\bibinfo {title}
  {Antiferromagnetism and d-wave superconductivity in cuprates: A cluster
  dynamical mean-field theory},}\ }\href {\doibase 10.1103/PhysRevB.62.R9283}
  {\bibfield  {journal} {\bibinfo  {journal} {Phys. Rev. B}\ }\textbf {\bibinfo
  {volume} {62}},\ \bibinfo {pages} {R9283--R9286} (\bibinfo {year}
  {2000})}\BibitemShut {NoStop}%
\bibitem [{\citenamefont {Georges}\ \emph {et~al.}(1996)\citenamefont
  {Georges}, \citenamefont {Kotliar}, \citenamefont {Krauth},\ and\
  \citenamefont {Rozenberg}}]{georges1996}%
  \BibitemOpen
  \bibfield  {author} {\bibinfo {author} {\bibfnamefont {A.}~\bibnamefont
  {Georges}}, \bibinfo {author} {\bibfnamefont {G.}~\bibnamefont {Kotliar}},
  \bibinfo {author} {\bibfnamefont {W.}~\bibnamefont {Krauth}}, \ and\ \bibinfo
  {author} {\bibfnamefont {M.~J.}\ \bibnamefont {Rozenberg}},\ }\bibfield
  {title} {\enquote {\bibinfo {title} {Dynamical mean-field theory of strongly
  correlated fermion systems and the limit of infinite dimensions},}\ }\href
  {\doibase 10.1103/RevModPhys.68.13} {\bibfield  {journal} {\bibinfo
  {journal} {Rev. Mod. Phys.}\ }\textbf {\bibinfo {volume} {68}},\ \bibinfo
  {pages} {13--125} (\bibinfo {year} {1996})}\BibitemShut {NoStop}%
\bibitem [{\citenamefont {Aron}\ \emph {et~al.}(2013)\citenamefont {Aron},
  \citenamefont {Weber},\ and\ \citenamefont {Kotliar}}]{aron2013}%
  \BibitemOpen
  \bibfield  {author} {\bibinfo {author} {\bibfnamefont {C.}~\bibnamefont
  {Aron}}, \bibinfo {author} {\bibfnamefont {C.}~\bibnamefont {Weber}}, \ and\
  \bibinfo {author} {\bibfnamefont {G.}~\bibnamefont {Kotliar}},\ }\bibfield
  {title} {\enquote {\bibinfo {title} {Impurity model for non-equilibrium
  steady states},}\ }\href {\doibase 10.1103/PhysRevB.87.125113} {\bibfield
  {journal} {\bibinfo  {journal} {Phys. Rev. B}\ }\textbf {\bibinfo {volume}
  {87}},\ \bibinfo {pages} {125113} (\bibinfo {year} {2013})}\BibitemShut
  {NoStop}%
\bibitem [{\citenamefont {Aoki}\ \emph {et~al.}(2014)\citenamefont {Aoki},
  \citenamefont {Tsuji}, \citenamefont {Eckstein}, \citenamefont {Kollar},
  \citenamefont {Oka},\ and\ \citenamefont {Werner}}]{aoki2014}%
  \BibitemOpen
  \bibfield  {author} {\bibinfo {author} {\bibfnamefont {H.}~\bibnamefont
  {Aoki}}, \bibinfo {author} {\bibfnamefont {N.}~\bibnamefont {Tsuji}},
  \bibinfo {author} {\bibfnamefont {M.}~\bibnamefont {Eckstein}}, \bibinfo
  {author} {\bibfnamefont {M.}~\bibnamefont {Kollar}}, \bibinfo {author}
  {\bibfnamefont {T.}~\bibnamefont {Oka}}, \ and\ \bibinfo {author}
  {\bibfnamefont {P.}~\bibnamefont {Werner}},\ }\bibfield  {title} {\enquote
  {\bibinfo {title} {Nonequilibrium dynamical mean-field theory and its
  applications},}\ }\href {\doibase 10.1103/RevModPhys.86.779} {\bibfield
  {journal} {\bibinfo  {journal} {Rev. Mod. Phys.}\ }\textbf {\bibinfo {volume}
  {86}},\ \bibinfo {pages} {779--837} (\bibinfo {year} {2014})}\BibitemShut
  {NoStop}%
\bibitem [{\citenamefont {Li}\ \emph {et~al.}(2015)\citenamefont {Li},
  \citenamefont {Aron}, \citenamefont {Kotliar},\ and\ \citenamefont
  {Han}}]{li2015}%
  \BibitemOpen
  \bibfield  {author} {\bibinfo {author} {\bibfnamefont {J.}~\bibnamefont
  {Li}}, \bibinfo {author} {\bibfnamefont {C.}~\bibnamefont {Aron}}, \bibinfo
  {author} {\bibfnamefont {G.}~\bibnamefont {Kotliar}}, \ and\ \bibinfo
  {author} {\bibfnamefont {J.~E.}\ \bibnamefont {Han}},\ }\bibfield  {title}
  {\enquote {\bibinfo {title} {Electric-field-driven resistive switching in the
  dissipative hubbard model},}\ }\href {\doibase
  10.1103/PhysRevLett.114.226403} {\bibfield  {journal} {\bibinfo  {journal}
  {Phys. Rev. Lett.}\ }\textbf {\bibinfo {volume} {114}},\ \bibinfo {pages}
  {226403} (\bibinfo {year} {2015})}\BibitemShut {NoStop}%
\bibitem [{\citenamefont {Ayral}\ \emph {et~al.}(2017)\citenamefont {Ayral},
  \citenamefont {Biermann}, \citenamefont {Werner},\ and\ \citenamefont
  {Boehnke}}]{ayral2017}%
  \BibitemOpen
  \bibfield  {author} {\bibinfo {author} {\bibfnamefont {T.}~\bibnamefont
  {Ayral}}, \bibinfo {author} {\bibfnamefont {S.}~\bibnamefont {Biermann}},
  \bibinfo {author} {\bibfnamefont {P.}~\bibnamefont {Werner}}, \ and\ \bibinfo
  {author} {\bibfnamefont {L.}~\bibnamefont {Boehnke}},\ }\bibfield  {title}
  {\enquote {\bibinfo {title} {Influence of fock exchange in combined many-body
  perturbation and dynamical mean field theory},}\ }\href {\doibase
  10.1103/PhysRevB.95.245130} {\bibfield  {journal} {\bibinfo  {journal} {Phys.
  Rev. B}\ }\textbf {\bibinfo {volume} {95}},\ \bibinfo {pages} {245130}
  (\bibinfo {year} {2017})}\BibitemShut {NoStop}%
\bibitem [{\citenamefont {Singer}\ \emph {et~al.}(2005)\citenamefont {Singer},
  \citenamefont {Reetz-Lamour}, \citenamefont {Amthor}, \citenamefont
  {Folling}, \citenamefont {Tscherneck},\ and\ \citenamefont
  {Weidem\"uller}}]{singer2005}%
  \BibitemOpen
  \bibfield  {author} {\bibinfo {author} {\bibfnamefont {K.}~\bibnamefont
  {Singer}}, \bibinfo {author} {\bibfnamefont {M.}~\bibnamefont
  {Reetz-Lamour}}, \bibinfo {author} {\bibfnamefont {T.}~\bibnamefont
  {Amthor}}, \bibinfo {author} {\bibfnamefont {S.}~\bibnamefont {Folling}},
  \bibinfo {author} {\bibfnamefont {M.}~\bibnamefont {Tscherneck}}, \ and\
  \bibinfo {author} {\bibfnamefont {M.}~\bibnamefont {Weidem\"uller}},\
  }\bibfield  {title} {\enquote {\bibinfo {title} {Spectroscopy of an ultracold
  rydberg gas and signatures of rydberg-rydberg interactions},}\ }\href
  {\doibase 10.1088/0953-4075/38/2/023} {\bibfield  {journal} {\bibinfo
  {journal} {J. Phys. B}\ }\textbf {\bibinfo {volume} {38}},\ \bibinfo {pages}
  {S321} (\bibinfo {year} {2005})}\BibitemShut {NoStop}%
\bibitem [{\citenamefont {Saffman}\ \emph {et~al.}(2010)\citenamefont
  {Saffman}, \citenamefont {Walker},\ and\ \citenamefont
  {M\o{}lmer}}]{saffman2010}%
  \BibitemOpen
  \bibfield  {author} {\bibinfo {author} {\bibfnamefont {M.}~\bibnamefont
  {Saffman}}, \bibinfo {author} {\bibfnamefont {T.~G.}\ \bibnamefont {Walker}},
  \ and\ \bibinfo {author} {\bibfnamefont {K.}~\bibnamefont {M\o{}lmer}},\
  }\bibfield  {title} {\enquote {\bibinfo {title} {Quantum information with
  rydberg atoms},}\ }\href {\doibase 10.1103/RevModPhys.82.2313} {\bibfield
  {journal} {\bibinfo  {journal} {Rev. Mod. Phys.}\ }\textbf {\bibinfo {volume}
  {82}},\ \bibinfo {pages} {2313--2363} (\bibinfo {year} {2010})}\BibitemShut
  {NoStop}%
\bibitem [{\citenamefont {Cohen-Tannoudji}\ \emph {et~al.}(1998)\citenamefont
  {Cohen-Tannoudji}, \citenamefont {Dupont-Roc},\ and\ \citenamefont
  {Grynberg}}]{cohen1998}%
  \BibitemOpen
  \bibfield  {author} {\bibinfo {author} {\bibfnamefont {C.}~\bibnamefont
  {Cohen-Tannoudji}}, \bibinfo {author} {\bibfnamefont {J.}~\bibnamefont
  {Dupont-Roc}}, \ and\ \bibinfo {author} {\bibfnamefont {G.}~\bibnamefont
  {Grynberg}},\ }\href@noop {} {\emph {\bibinfo {title} {Atom-Photon
  Interactions: Basic Processes and Applications}}}\ (\bibinfo  {publisher}
  {Wiley, New York},\ \bibinfo {year} {1998})\BibitemShut {NoStop}%
\bibitem [{\citenamefont {Balewski}\ \emph {et~al.}(2014)\citenamefont
  {Balewski}, \citenamefont {Krupp}, \citenamefont {Gaj}, \citenamefont
  {Hofferberth}, \citenamefont {L\"ow},\ and\ \citenamefont
  {Pfau}}]{balewski2014}%
  \BibitemOpen
  \bibfield  {author} {\bibinfo {author} {\bibfnamefont {J.~B.}\ \bibnamefont
  {Balewski}}, \bibinfo {author} {\bibfnamefont {A.~T.}\ \bibnamefont {Krupp}},
  \bibinfo {author} {\bibfnamefont {A.}~\bibnamefont {Gaj}}, \bibinfo {author}
  {\bibfnamefont {S.}~\bibnamefont {Hofferberth}}, \bibinfo {author}
  {\bibfnamefont {R.}~\bibnamefont {L\"ow}}, \ and\ \bibinfo {author}
  {\bibfnamefont {T.}~\bibnamefont {Pfau}},\ }\bibfield  {title} {\enquote
  {\bibinfo {title} {Rydberg dressing: understanding of collective many-body
  effects and implications for experiments},}\ }\href
  {https://doi.org/10.1088/1367-2630/16/6/063012} {\bibfield  {journal}
  {\bibinfo  {journal} {New J. Phys.}\ }\textbf {\bibinfo {volume} {16}},\
  \bibinfo {pages} {063012} (\bibinfo {year} {2014})}\BibitemShut {NoStop}%
\bibitem [{\citenamefont {Henkel}\ \emph {et~al.}(2010)\citenamefont {Henkel},
  \citenamefont {Nath},\ and\ \citenamefont {Pohl}}]{henkel2010}%
  \BibitemOpen
  \bibfield  {author} {\bibinfo {author} {\bibfnamefont {N.}~\bibnamefont
  {Henkel}}, \bibinfo {author} {\bibfnamefont {R.}~\bibnamefont {Nath}}, \ and\
  \bibinfo {author} {\bibfnamefont {T.}~\bibnamefont {Pohl}},\ }\bibfield
  {title} {\enquote {\bibinfo {title} {Three-dimensional roton excitations and
  supersolid formation in rydberg-excited bose-einstein condensates},}\ }\href
  {\doibase 10.1103/PhysRevLett.104.195302} {\bibfield  {journal} {\bibinfo
  {journal} {Phys. Rev. Lett.}\ }\textbf {\bibinfo {volume} {104}},\ \bibinfo
  {pages} {195302} (\bibinfo {year} {2010})}\BibitemShut {NoStop}%
\bibitem [{\citenamefont {Viteau}\ \emph {et~al.}(2011)\citenamefont {Viteau},
  \citenamefont {Bason}, \citenamefont {Radogostowicz}, \citenamefont
  {Malossi}, \citenamefont {Ciampini}, \citenamefont {Morsch},\ and\
  \citenamefont {Arimondo}}]{viteau2011}%
  \BibitemOpen
  \bibfield  {author} {\bibinfo {author} {\bibfnamefont {M.}~\bibnamefont
  {Viteau}}, \bibinfo {author} {\bibfnamefont {M.~G.}\ \bibnamefont {Bason}},
  \bibinfo {author} {\bibfnamefont {J.}~\bibnamefont {Radogostowicz}}, \bibinfo
  {author} {\bibfnamefont {N.}~\bibnamefont {Malossi}}, \bibinfo {author}
  {\bibfnamefont {D.}~\bibnamefont {Ciampini}}, \bibinfo {author}
  {\bibfnamefont {O.}~\bibnamefont {Morsch}}, \ and\ \bibinfo {author}
  {\bibfnamefont {E.}~\bibnamefont {Arimondo}},\ }\bibfield  {title} {\enquote
  {\bibinfo {title} {Rydberg excitations in bose-einstein condensates in
  quasi-one-dimensional potentials and optical lattices},}\ }\href {\doibase
  10.1103/PhysRevLett.107.060402} {\bibfield  {journal} {\bibinfo  {journal}
  {Phys. Rev. Lett.}\ }\textbf {\bibinfo {volume} {107}},\ \bibinfo {pages}
  {060402} (\bibinfo {year} {2011})}\BibitemShut {NoStop}%
\bibitem [{\citenamefont {L\"ow}\ \emph {et~al.}(2012)\citenamefont {L\"ow},
  \citenamefont {Weimer}, \citenamefont {Nipper}, \citenamefont {Balewski},
  \citenamefont {Butscher}, \citenamefont {B\"uchler},\ and\ \citenamefont
  {Pfau}}]{loew2012}%
  \BibitemOpen
  \bibfield  {author} {\bibinfo {author} {\bibfnamefont {R.}~\bibnamefont
  {L\"ow}}, \bibinfo {author} {\bibfnamefont {H.}~\bibnamefont {Weimer}},
  \bibinfo {author} {\bibfnamefont {J.}~\bibnamefont {Nipper}}, \bibinfo
  {author} {\bibfnamefont {J.~B.}\ \bibnamefont {Balewski}}, \bibinfo {author}
  {\bibfnamefont {B.}~\bibnamefont {Butscher}}, \bibinfo {author}
  {\bibfnamefont {H.~P.}\ \bibnamefont {B\"uchler}}, \ and\ \bibinfo {author}
  {\bibfnamefont {T.}~\bibnamefont {Pfau}},\ }\bibfield  {title} {\enquote
  {\bibinfo {title} {An experimental and theoretical guide to strongly
  interacting rydberg gases},}\ }\href
  {http://stacks.iop.org/0953-4075/45/i=11/a=113001} {\bibfield  {journal}
  {\bibinfo  {journal} {J.~Phys.~B}\ }\textbf {\bibinfo {volume} {45}},\
  \bibinfo {pages} {113001} (\bibinfo {year} {2012})}\BibitemShut {NoStop}%
\bibitem [{\citenamefont {DeSalvo}\ \emph {et~al.}(2016)\citenamefont
  {DeSalvo}, \citenamefont {Aman}, \citenamefont {Gaul}, \citenamefont {Pohl},
  \citenamefont {Yoshida}, \citenamefont {Burgd\"orfer}, \citenamefont
  {Hazzard}, \citenamefont {Dunning},\ and\ \citenamefont
  {Killian}}]{desalvo2016}%
  \BibitemOpen
  \bibfield  {author} {\bibinfo {author} {\bibfnamefont {B.~J.}\ \bibnamefont
  {DeSalvo}}, \bibinfo {author} {\bibfnamefont {J.~A.}\ \bibnamefont {Aman}},
  \bibinfo {author} {\bibfnamefont {C.}~\bibnamefont {Gaul}}, \bibinfo {author}
  {\bibfnamefont {T.}~\bibnamefont {Pohl}}, \bibinfo {author} {\bibfnamefont
  {S.}~\bibnamefont {Yoshida}}, \bibinfo {author} {\bibfnamefont
  {J.}~\bibnamefont {Burgd\"orfer}}, \bibinfo {author} {\bibfnamefont
  {K.~R.~A.}\ \bibnamefont {Hazzard}}, \bibinfo {author} {\bibfnamefont
  {F.~B.}\ \bibnamefont {Dunning}}, \ and\ \bibinfo {author} {\bibfnamefont
  {T.~C.}\ \bibnamefont {Killian}},\ }\bibfield  {title} {\enquote {\bibinfo
  {title} {Rydberg-blockade effects in autler-townes spectra of ultracold
  strontium},}\ }\href {\doibase 10.1103/PhysRevA.93.022709} {\bibfield
  {journal} {\bibinfo  {journal} {Phys. Rev. A}\ }\textbf {\bibinfo {volume}
  {93}},\ \bibinfo {pages} {022709} (\bibinfo {year} {2016})}\BibitemShut
  {NoStop}%
\bibitem [{\citenamefont {Davies}(1978)}]{davies1978}%
  \BibitemOpen
  \bibfield  {author} {\bibinfo {author} {\bibfnamefont {E.~B.}\ \bibnamefont
  {Davies}},\ }\bibfield  {title} {\enquote {\bibinfo {title} {A model of heat
  conduction},}\ }\href {\doibase 10.1007/BF01014307} {\bibfield  {journal}
  {\bibinfo  {journal} {Journal of Statistical Physics}\ }\textbf {\bibinfo
  {volume} {18}},\ \bibinfo {pages} {161--170} (\bibinfo {year}
  {1978})}\BibitemShut {NoStop}%
\bibitem [{\citenamefont {B\"uttiker}(1985)}]{buttiker1985}%
  \BibitemOpen
  \bibfield  {author} {\bibinfo {author} {\bibfnamefont {M.}~\bibnamefont
  {B\"uttiker}},\ }\bibfield  {title} {\enquote {\bibinfo {title} {Small
  normal-metal loop coupled to an electron reservoir},}\ }\href {\doibase
  10.1103/PhysRevB.32.1846} {\bibfield  {journal} {\bibinfo  {journal} {Phys.
  Rev. B}\ }\textbf {\bibinfo {volume} {32}},\ \bibinfo {pages} {1846--1849}
  (\bibinfo {year} {1985})}\BibitemShut {NoStop}%
\bibitem [{\citenamefont {B\"uttiker}(1986)}]{buttiker1986}%
  \BibitemOpen
  \bibfield  {author} {\bibinfo {author} {\bibfnamefont {M.}~\bibnamefont
  {B\"uttiker}},\ }\bibfield  {title} {\enquote {\bibinfo {title} {Role of
  quantum coherence in series resistors},}\ }\href {\doibase
  10.1103/PhysRevB.33.3020} {\bibfield  {journal} {\bibinfo  {journal} {Phys.
  Rev. B}\ }\textbf {\bibinfo {volume} {33}},\ \bibinfo {pages} {3020--3026}
  (\bibinfo {year} {1986})}\BibitemShut {NoStop}%
\bibitem [{\citenamefont {Tsuji}(2010)}]{tsuji2010}%
  \BibitemOpen
  \bibfield  {author} {\bibinfo {author} {\bibfnamefont {N.}~\bibnamefont
  {Tsuji}},\ }\emph {\bibinfo {title} {Theoretical Study of Nonequilibrium
  Correlated Fermions Driven by ac Fields}},\ \href@noop {} {Ph.D. thesis},\
  \bibinfo  {school} {University of Tokyo} (\bibinfo {year} {2010})\BibitemShut
  {NoStop}%
\bibitem [{\citenamefont {Qin}\ and\ \citenamefont
  {Hofstetter}(2018)}]{qin2018}%
  \BibitemOpen
  \bibfield  {author} {\bibinfo {author} {\bibfnamefont {T.}~\bibnamefont
  {Qin}}\ and\ \bibinfo {author} {\bibfnamefont {W.}~\bibnamefont
  {Hofstetter}},\ }\bibfield  {title} {\enquote {\bibinfo {title}
  {Nonequilibrium steady states and resonant tunneling in time-periodically
  driven systems with interactions},}\ }\href {\doibase
  10.1103/PhysRevB.97.125115} {\bibfield  {journal} {\bibinfo  {journal} {Phys.
  Rev. B}\ }\textbf {\bibinfo {volume} {97}},\ \bibinfo {pages} {125115}
  (\bibinfo {year} {2018})}\BibitemShut {NoStop}%
\bibitem [{\citenamefont {Yan}\ \emph {et~al.}(2018)\citenamefont {Yan},
  \citenamefont {Pollet}, \citenamefont {Lou}, \citenamefont {Wang},
  \citenamefont {Chen},\ and\ \citenamefont {Cai}}]{yan2018}%
  \BibitemOpen
  \bibfield  {author} {\bibinfo {author} {\bibfnamefont {Z.}~\bibnamefont
  {Yan}}, \bibinfo {author} {\bibfnamefont {L.}~\bibnamefont {Pollet}},
  \bibinfo {author} {\bibfnamefont {J.}~\bibnamefont {Lou}}, \bibinfo {author}
  {\bibfnamefont {X.}~\bibnamefont {Wang}}, \bibinfo {author} {\bibfnamefont
  {Y.}~\bibnamefont {Chen}}, \ and\ \bibinfo {author} {\bibfnamefont
  {Z.}~\bibnamefont {Cai}},\ }\bibfield  {title} {\enquote {\bibinfo {title}
  {Interacting lattice systems with quantum dissipation: A quantum monte carlo
  study},}\ }\href {\doibase 10.1103/PhysRevB.97.035148} {\bibfield  {journal}
  {\bibinfo  {journal} {Phys. Rev. B}\ }\textbf {\bibinfo {volume} {97}},\
  \bibinfo {pages} {035148} (\bibinfo {year} {2018})}\BibitemShut {NoStop}%
\bibitem [{\citenamefont {Peronaci}\ \emph {et~al.}(2018)\citenamefont
  {Peronaci}, \citenamefont {Schir\'o},\ and\ \citenamefont
  {Parcollet}}]{peronaci2018}%
  \BibitemOpen
  \bibfield  {author} {\bibinfo {author} {\bibfnamefont {F.}~\bibnamefont
  {Peronaci}}, \bibinfo {author} {\bibfnamefont {M.}~\bibnamefont {Schir\'o}},
  \ and\ \bibinfo {author} {\bibfnamefont {O.}~\bibnamefont {Parcollet}},\
  }\bibfield  {title} {\enquote {\bibinfo {title} {Resonant thermalization of
  periodically driven strongly correlated electrons},}\ }\href {\doibase
  10.1103/PhysRevLett.120.197601} {\bibfield  {journal} {\bibinfo  {journal}
  {Phys. Rev. Lett.}\ }\textbf {\bibinfo {volume} {120}},\ \bibinfo {pages}
  {197601} (\bibinfo {year} {2018})}\BibitemShut {NoStop}%
\bibitem [{\citenamefont {Chong}\ \emph {et~al.}(2018)\citenamefont {Chong},
  \citenamefont {Kim}, \citenamefont {Kim}, \citenamefont {Yoon}, \citenamefont
  {Kang},\ and\ \citenamefont {An}}]{chong2018}%
  \BibitemOpen
  \bibfield  {author} {\bibinfo {author} {\bibfnamefont {K.~O.}\ \bibnamefont
  {Chong}}, \bibinfo {author} {\bibfnamefont {J.-R.}\ \bibnamefont {Kim}},
  \bibinfo {author} {\bibfnamefont {J.}~\bibnamefont {Kim}}, \bibinfo {author}
  {\bibfnamefont {S.}~\bibnamefont {Yoon}}, \bibinfo {author} {\bibfnamefont
  {S.}~\bibnamefont {Kang}}, \ and\ \bibinfo {author} {\bibfnamefont
  {K.}~\bibnamefont {An}},\ }\bibfield  {title} {\enquote {\bibinfo {title}
  {Observation of a non-equilibrium steady state of cold atoms in a moving
  optical lattice},}\ }\href {\doibase 10.1038/s42005-018-0024-5} {\bibfield
  {journal} {\bibinfo  {journal} {Communications Physics}\ }\textbf {\bibinfo
  {volume} {1}},\ \bibinfo {pages} {25} (\bibinfo {year} {2018})}\BibitemShut
  {NoStop}%
\bibitem [{Note1()}]{Note1}%
  \BibitemOpen
  \bibinfo {note} {If we neglect the Keldysh part in the decoupled system it is
  essential that the full system can thermalize either through interaction or
  through coupling to thermal bath.}\BibitemShut {Stop}%
\bibitem [{Note2()}]{Note2}%
  \BibitemOpen
  \bibinfo {note} {It includes effects of non-local interaction treated on the
  mean-field level}\BibitemShut {NoStop}%
\bibitem [{\citenamefont {Dzhioev}\ and\ \citenamefont
  {Kosov}(2011{\natexlab{b}})}]{dzhioev2011super}%
  \BibitemOpen
  \bibfield  {author} {\bibinfo {author} {\bibfnamefont {A.~A.}\ \bibnamefont
  {Dzhioev}}\ and\ \bibinfo {author} {\bibfnamefont {D.~S.}\ \bibnamefont
  {Kosov}},\ }\bibfield  {title} {\enquote {\bibinfo {title} {Super-fermion
  representation of quantum kinetic equations for the electron transport
  problem},}\ }\href {\doibase 10.1063/1.3548065} {\bibfield  {journal}
  {\bibinfo  {journal} {The Journal of Chemical Physics}\ }\textbf {\bibinfo
  {volume} {134}},\ \bibinfo {pages} {044121} (\bibinfo {year}
  {2011}{\natexlab{b}})},\ \Eprint
  {http://arxiv.org/abs/https://doi.org/10.1063/1.3548065}
  {https://doi.org/10.1063/1.3548065} \BibitemShut {NoStop}%
\bibitem [{\citenamefont {Sorantin}\ \emph {et~al.}(2018)\citenamefont
  {Sorantin}, \citenamefont {Fugger}, \citenamefont {Dorda}, \citenamefont
  {von~der Linden},\ and\ \citenamefont {Arrigoni}}]{sorantin2018}%
  \BibitemOpen
  \bibfield  {author} {\bibinfo {author} {\bibfnamefont {M.~E.}\ \bibnamefont
  {Sorantin}}, \bibinfo {author} {\bibfnamefont {D.~M.}\ \bibnamefont
  {Fugger}}, \bibinfo {author} {\bibfnamefont {A.}~\bibnamefont {Dorda}},
  \bibinfo {author} {\bibfnamefont {W.}~\bibnamefont {von~der Linden}}, \ and\
  \bibinfo {author} {\bibfnamefont {E.}~\bibnamefont {Arrigoni}},\ }\bibfield
  {title} {\enquote {\bibinfo {title} {Auxiliary master equation approach
  within stochastic wave functions: Application to the interacting resonant
  level model},}\ }\href@noop {} {\bibfield  {journal} {\bibinfo  {journal} {to
  be published}\ } (\bibinfo {year} {2018})}\BibitemShut {NoStop}%
\bibitem [{\citenamefont {Zhang}\ \emph {et~al.}(2012)\citenamefont {Zhang},
  \citenamefont {Lo}, \citenamefont {Xiong}, \citenamefont {Tu},\ and\
  \citenamefont {Nori}}]{zhang2012}%
  \BibitemOpen
  \bibfield  {author} {\bibinfo {author} {\bibfnamefont {W.-M.}\ \bibnamefont
  {Zhang}}, \bibinfo {author} {\bibfnamefont {P.-Y.}\ \bibnamefont {Lo}},
  \bibinfo {author} {\bibfnamefont {H.-N.}\ \bibnamefont {Xiong}}, \bibinfo
  {author} {\bibfnamefont {M.~W.-Y.}\ \bibnamefont {Tu}}, \ and\ \bibinfo
  {author} {\bibfnamefont {F.}~\bibnamefont {Nori}},\ }\bibfield  {title}
  {\enquote {\bibinfo {title} {General non-markovian dynamics of open quantum
  systems},}\ }\href {\doibase 10.1103/PhysRevLett.109.170402} {\bibfield
  {journal} {\bibinfo  {journal} {Phys. Rev. Lett.}\ }\textbf {\bibinfo
  {volume} {109}},\ \bibinfo {pages} {170402} (\bibinfo {year}
  {2012})}\BibitemShut {NoStop}%
\bibitem [{\citenamefont {Hafez-Torbati}\ and\ \citenamefont
  {Hofstetter}(2018)}]{hafez-torbati2018}%
  \BibitemOpen
  \bibfield  {author} {\bibinfo {author} {\bibfnamefont {H.}~\bibnamefont
  {Hafez-Torbati}}\ and\ \bibinfo {author} {\bibfnamefont {W.}~\bibnamefont
  {Hofstetter}},\ }\bibfield  {title} {\enquote {\bibinfo {title} {Artificial
  su(3) spin-orbit coupling and exotic mott insulators},}\ }\href
  {https://arxiv.org/abs/1809.05704} {\bibfield  {journal} {\bibinfo  {journal}
  {arXiv}\ ,\ \bibinfo {pages} {1809.05704}} (\bibinfo {year}
  {2018})}\BibitemShut {NoStop}%
\bibitem [{\citenamefont {Blaizot}\ and\ \citenamefont
  {Ripka}(1986)}]{blaizot1986}%
  \BibitemOpen
  \bibfield  {author} {\bibinfo {author} {\bibfnamefont {J.-P.}\ \bibnamefont
  {Blaizot}}\ and\ \bibinfo {author} {\bibfnamefont {G.}~\bibnamefont
  {Ripka}},\ }\href@noop {} {\emph {\bibinfo {title} {Quantum Theory of Finite
  Systems}}}\ (\bibinfo  {publisher} {Cambridge, MA},\ \bibinfo {year}
  {1986})\BibitemShut {NoStop}%
\bibitem [{\citenamefont {Bhongale}\ \emph {et~al.}(2012)\citenamefont
  {Bhongale}, \citenamefont {Mathey}, \citenamefont {Tsai}, \citenamefont
  {Clark},\ and\ \citenamefont {Zhao}}]{bhongale2012}%
  \BibitemOpen
  \bibfield  {author} {\bibinfo {author} {\bibfnamefont {S.~G.}\ \bibnamefont
  {Bhongale}}, \bibinfo {author} {\bibfnamefont {L.}~\bibnamefont {Mathey}},
  \bibinfo {author} {\bibfnamefont {S.-W.}\ \bibnamefont {Tsai}}, \bibinfo
  {author} {\bibfnamefont {C.~W.}\ \bibnamefont {Clark}}, \ and\ \bibinfo
  {author} {\bibfnamefont {E.}~\bibnamefont {Zhao}},\ }\bibfield  {title}
  {\enquote {\bibinfo {title} {Bond order solid of two-dimensional dipolar
  fermions},}\ }\href {\doibase 10.1103/PhysRevLett.108.145301} {\bibfield
  {journal} {\bibinfo  {journal} {Phys. Rev. Lett.}\ }\textbf {\bibinfo
  {volume} {108}},\ \bibinfo {pages} {145301} (\bibinfo {year}
  {2012})}\BibitemShut {NoStop}%
\bibitem [{\citenamefont {Bhongale}\ \emph {et~al.}(2013)\citenamefont
  {Bhongale}, \citenamefont {Mathey}, \citenamefont {Tsai}, \citenamefont
  {Clark},\ and\ \citenamefont {Zhao}}]{bhongale2013}%
  \BibitemOpen
  \bibfield  {author} {\bibinfo {author} {\bibfnamefont {S.~G.}\ \bibnamefont
  {Bhongale}}, \bibinfo {author} {\bibfnamefont {L.}~\bibnamefont {Mathey}},
  \bibinfo {author} {\bibfnamefont {S.-W.}\ \bibnamefont {Tsai}}, \bibinfo
  {author} {\bibfnamefont {C.~W.}\ \bibnamefont {Clark}}, \ and\ \bibinfo
  {author} {\bibfnamefont {E.}~\bibnamefont {Zhao}},\ }\bibfield  {title}
  {\enquote {\bibinfo {title} {Unconventional spin-density waves in dipolar
  fermi gases},}\ }\href {\doibase 10.1103/PhysRevA.87.043604} {\bibfield
  {journal} {\bibinfo  {journal} {Phys. Rev. A}\ }\textbf {\bibinfo {volume}
  {87}},\ \bibinfo {pages} {043604} (\bibinfo {year} {2013})}\BibitemShut
  {NoStop}%
\bibitem [{\citenamefont {Titvinidze}\ \emph {et~al.}(2017)\citenamefont
  {Titvinidze}, \citenamefont {Dorda}, \citenamefont {von~der Linden},\ and\
  \citenamefont {Arrigoni}}]{titvinidze2017}%
  \BibitemOpen
  \bibfield  {author} {\bibinfo {author} {\bibfnamefont {Irakli}\ \bibnamefont
  {Titvinidze}}, \bibinfo {author} {\bibfnamefont {Antonius}\ \bibnamefont
  {Dorda}}, \bibinfo {author} {\bibfnamefont {Wolfgang}\ \bibnamefont {von~der
  Linden}}, \ and\ \bibinfo {author} {\bibfnamefont {Enrico}\ \bibnamefont
  {Arrigoni}},\ }\bibfield  {title} {\enquote {\bibinfo {title} {Thermoelectric
  properties of a strongly correlated layer},}\ }\href {\doibase
  10.1103/PhysRevB.96.115104} {\bibfield  {journal} {\bibinfo  {journal} {Phys.
  Rev. B}\ }\textbf {\bibinfo {volume} {96}},\ \bibinfo {pages} {115104}
  (\bibinfo {year} {2017})}\BibitemShut {NoStop}%
\bibitem [{\citenamefont {Beterov}\ \emph {et~al.}(2009)\citenamefont
  {Beterov}, \citenamefont {Ryabtsev}, \citenamefont {Tretyakov},\ and\
  \citenamefont {Entin}}]{beterov2009}%
  \BibitemOpen
  \bibfield  {author} {\bibinfo {author} {\bibfnamefont {I.~I.}\ \bibnamefont
  {Beterov}}, \bibinfo {author} {\bibfnamefont {I.~I.}\ \bibnamefont
  {Ryabtsev}}, \bibinfo {author} {\bibfnamefont {D.~B.}\ \bibnamefont
  {Tretyakov}}, \ and\ \bibinfo {author} {\bibfnamefont {V.~M.}\ \bibnamefont
  {Entin}},\ }\bibfield  {title} {\enquote {\bibinfo {title} {Quasiclassical
  calculations of blackbody-radiation-induced depopulation rates and effective
  lifetimes of rydberg $ns$, $np$, and $nd$ alkali-metal atoms with
  $n\ensuremath{\le}80$},}\ }\href {\doibase 10.1103/PhysRevA.79.052504}
  {\bibfield  {journal} {\bibinfo  {journal} {Phys. Rev. A}\ }\textbf {\bibinfo
  {volume} {79}},\ \bibinfo {pages} {052504} (\bibinfo {year}
  {2009})}\BibitemShut {NoStop}%
\bibitem [{\citenamefont {Saha}\ \emph {et~al.}(2014)\citenamefont {Saha},
  \citenamefont {Sinha},\ and\ \citenamefont {Sengupta}}]{saha2014}%
  \BibitemOpen
  \bibfield  {author} {\bibinfo {author} {\bibfnamefont {K.}~\bibnamefont
  {Saha}}, \bibinfo {author} {\bibfnamefont {S.}~\bibnamefont {Sinha}}, \ and\
  \bibinfo {author} {\bibfnamefont {K.}~\bibnamefont {Sengupta}},\ }\bibfield
  {title} {\enquote {\bibinfo {title} {Phases and collective modes of rydberg
  atoms in an optical lattice},}\ }\href {\doibase 10.1103/PhysRevA.89.023618}
  {\bibfield  {journal} {\bibinfo  {journal} {Phys. Rev. A}\ }\textbf {\bibinfo
  {volume} {89}},\ \bibinfo {pages} {023618} (\bibinfo {year}
  {2014})}\BibitemShut {NoStop}%
\bibitem [{\citenamefont {Spohn}(1976)}]{spohn1976}%
  \BibitemOpen
  \bibfield  {author} {\bibinfo {author} {\bibfnamefont {H.}~\bibnamefont
  {Spohn}},\ }\bibfield  {title} {\enquote {\bibinfo {title} {Approach to
  equilibrium for completely positive dynamical semigroups of n-level
  systems},}\ }\href {\doibase https://doi.org/10.1016/0034-4877(76)90040-9}
  {\bibfield  {journal} {\bibinfo  {journal} {Reports on Mathematical Physics}\
  }\textbf {\bibinfo {volume} {10}},\ \bibinfo {pages} {189 -- 194} (\bibinfo
  {year} {1976})}\BibitemShut {NoStop}%
\bibitem [{\citenamefont {Spohn}(1977)}]{spohn1977}%
  \BibitemOpen
  \bibfield  {author} {\bibinfo {author} {\bibfnamefont {H.}~\bibnamefont
  {Spohn}},\ }\bibfield  {title} {\enquote {\bibinfo {title} {An algebraic
  condition for the approach to equilibrium of an open n-level system},}\
  }\href {\doibase 10.1007/BF00420668} {\bibfield  {journal} {\bibinfo
  {journal} {Letters in Mathematical Physics}\ }\textbf {\bibinfo {volume}
  {2}},\ \bibinfo {pages} {33--38} (\bibinfo {year} {1977})}\BibitemShut
  {NoStop}%
\end{thebibliography}%

\end{document}